\documentclass[aps,prd,preprint,floats,superscriptaddress,nofootinbib]{revtex4}
\usepackage[utf8]{inputenc}
\usepackage{amsmath,amssymb,bbm}
\usepackage{dsfont}
\usepackage{graphicx}
\usepackage{color,slashed}
\pdfoutput=1

\begin{document}
\title{\bf  Rare $B$ and $K$ decays in a scotogenic model}

\author{Chuan-Hung Chen}
\email[E-mail: ]{physchen@mail.ncku.edu.tw}
\affiliation{Department of Physics, National Cheng-Kung University, Tainan 70101, Taiwan}
\affiliation{Physics Division, National Center for Theoretical Sciences, Taipei 10617, Taiwan}

\author{Cheng-Wei Chiang}
\email[E-mail: ]{chengwei@phys.ntu.edu.tw}
\affiliation{Department of Physics and Center for Theoretical Physics, National Taiwan University, Taipei 10617, Taiwan}
\affiliation{Physics Division, National Center for Theoretical Sciences, Taipei 10617, Taiwan}

\date{\today}

\begin{abstract}

A scotogenic model can radiatively generate the observed neutrino mass, provide a dark matter candidate, and lead to rare lepton flavor-violating processes. We aim to extend the model to establish a potential connection to the quark flavor-related processes within the framework of scotogenesis, enhancing the unexpectedly large branching ratio (BR) of $B^+\to K^+ \nu \bar\nu$, observed by Belle II Collaboration. Meanwhile, the model can address tensions between some experimental measurements and standard model (SM) predictions in flavor physics, such as the muon $g-2$ excess and the higher BR of $B_s \to \mu^- \mu^+$. We introduce in the model the following dark particles: a neutral singlet Dirac-type lepton ($N$); two inert Higgs doublets ($\eta_{1,2}$), with one of which carrying a lepton number; a charged singlet dark scalar $(\chi^+)$, and a singlet vector-like up-type dark quark ($T$). The first two entities are responsible for the radiative neutrino mass, and $\chi^+$ couples to right-handed quarks and leptons and can resolve the tensions existing in muon $g-2$ and $B_s\to \mu^- \mu^+$. Furthermore, the BR of $B^+ \to K^+ \nu \bar\nu$ can be enhanced up to a factor of 2 compared to the SM prediction through the mediations of the dark $T$ and the charged scalars. In addition, we also study the impacts on the $K\to \pi \nu \bar\nu$ decays. 

\end{abstract}

\maketitle

\section{Introduction} \label{sec:intro}

Under an enormous number of experimental tests and with great success in most of them, the standard model (SM) has been established as a very good effective theory at and below the electroweak scale. However, certain empirical observations, such as the existence of neutrino mass and dark matter (DM), still await definitive resolutions. In addition, a long-standing issue in the anomalous magnetic dipole moment of muon (muon $g-2$), observed in BNL~\cite{Muong-2:2006rrc} and further confirmed by Fermilab experiments~\cite{Muong-2:2023cdq}, strongly hints at possibly a new interaction in the lepton sector.

Recently, the Belle II Collaboration with 362~fb$^{-1}$ of data has observed the first evidence of $B^+\to K^+ \nu\bar\nu$ decay, and the resulting branching ratio (BR) is reported as~\cite{Belle-II:2023esi}:
 \begin{align}
 {\cal B}(B^+ \to K^+ \nu \bar\nu) 
 = 
 [2.3 \pm 0.5 \, (\rm stat) \,^{+0.5}_{-0.4} \, (\rm syst)] \times 10^{-5}
 =
 (2.3\pm 0.7)\times 10^{-5}\,.
 \end{align}
 When combined with earlier results measured by BaBar~\cite{BaBar:2010oqg,BaBar:2013npw} and Belle~\cite{Belle:2013tnz,Belle:2017oht},  the weighted average is given by $ {\cal B}(B^+ \to K^+ \nu \bar\nu) =(1.3\pm 0.4)\times 10^{-5}$. Compared to the SM prediction of ${\cal B}(B^+\to K^+ \nu \bar\nu)^{\rm SM}=(4.92\pm 0.30) \times 10^{-6}$~\cite{Buras:2022qip}, the current data shows a $2.7\sigma$ deviation. This difference hints at the possibility of some peculiar interactions, predominantly manifesting in the $b\to s \nu \bar\nu$ or $b\to s + \text{invisible}$ transitions~\cite{Buras:2014fpa,Browder:2021hbl,Asadi:2023ucx,Athron:2023hmz,Bause:2023mfe,Allwicher:2023syp,Felkl:2023ayn,Dreiner:2023cms,Amhis:2023mpj,He:2023bnk,Chen:2023wpb,Datta:2023iln,Altmannshofer:2023hkn,Ho:2024cwk,Chen:2024jlj,Hou:2024vyw}, rather than in $b\to s \ell^- \ell^+$ that is subject to strict constraints from $B\to X_s \ell^- \ell^+$ and $B_s \to \mu^- \mu^+$ decays.

A primary motivation of this study is to extend the existing scotogenic model~\cite{Ma:2006km} to enhance the $d_i \to d_j \nu \bar \nu$ decays and muon $g-2$ while simultaneously explaining the observed neutrino data and dark matter relic density.  To radiatively generate Majorana neutrino mass in a scotogenic model, lepton number-violating couplings are essential. Such violation of the lepton number could stem from the leptonic right-handed Majorana neutrino mass term, like in the type-I seesaw mechanism~\cite{Minkowski:1977sc,Yanagida:1979as,Gell-Mann:1979vob,Mohapatra:1979ia}.  However, to prevent the mass scale of the introduced Majorana fermion from reaching an undetectable energy scale when the Yukawa couplings are of $O(m_\tau/v)$, either the Majorana fermion does not carry a lepton number~\cite{Ma:2006km} or a Dirac-type neutral fermion should be used instead~\cite{Chen:2022gmk}.  In other words, the new heavy fermion mass term is unrelated to or retains the lepton number conservation.  Therefore, within the framework of scotogenesis, the lepton number symmetry should be violated in some scalar coupling, for instance, a non-Hermitian quartic term, where the involved exotic scalar field carries a dark charge and is assigned a lepton number as proposed in the Ma model~\cite{Ma:2006km}. Since the lepton number symmetry will be restored when the scalar coupling, which violates the lepton number, approaches zero, it can be considered as a technical naturalness if the scalar coupling is small~\cite{tHooft:1979rat}.

If we focus on leptonic processes, one Higgs doublet, which carries a dark charge and lepton number, associated with $2$ or $3$ dark right-handed Majorana fermions in the Ma model can successfully account for the neutrino data observed from the neutrino oscillation experiments~\cite{Ma:2006km}. For simplicity, we will refer to a scalar carrying a dark charge as an inert scalar.  As a result, this model has implications for rare leptonic decays, such as $\mu\to e \gamma$, $\mu \to 3 e$, $\mu-e$ conversion in nucleus processes~\cite{Toma:2013zsa,Chen:2020ptg}, and DM candidate~\cite{Ma:2006km}.  However, due to the absence of the lepton quantum number in quarks, the lepton number-carrying Higgs doublet in the Ma model has no interactions with the quarks. This remains true even with the introduction of a new heavy quark carrying a dark charge; otherwise, the lepton number violation will occur in the quark Yukawa couplings, leading to the breakdown of $R_p$-parity, defined as $R_p=(-1)^{3B+L+2S}$, where $B$, $L$, and $S$ denote the baryon, lepton, and spin quantum numbers of a particle.

To incorporate the effects responsible for the loop-induced neutrino mass into the rare flavor-changing neutral-current (FCNC) $B$ and $K$ decay processes, a suitable extension of the Ma model is called for. We, therefore, aim to identify a minimal extension that not only addresses the issues of neutrino mass, DM relic density, and muon $g-2$ but also significantly enhances the BRs in the $d_i \to d_j \nu \bar\nu$ decays.

We find that the essential part to achieve our goal is the introduction of an inert Higgs doublet in the absence of the lepton number association. To preserve the lepton number conservation in the Yukawa sector, it is imperative to replace the Majorana-type neutral fermion used in the Ma model with a vector-like Dirac-type neutral lepton. Furthermore, to establish the connection between the SM quarks and the particles within the dark sector through the non-leptonic inert Higgs, it is necessary to introduce a new quark with an appropriate dark charge. For the sake of gauge anomaly-free conditions, the minimal choice of the new dark quark is an $SU(2)_L$ singlet vector-like up-type quark.

Since the non-leptonic inert Higgs is an $SU(2)_L$ doublet, the Yukawa couplings only involve the left-handed SM quarks. In other words, when $d_i\to d_j \nu \bar\nu $ are enhanced, the effects contributing to $B_s \to \mu^- \mu^+$ align with the SM contribution and could result in constructive interference and pushing its BR above the current experimental data. To address the constraint arising from $B_s\to \mu^- \mu^+$, a right-handed quark current used to cancel the effect from the left-handed current in the $b\to s$ transition becomes helpful. What is more is that in fact, the SM predicts ${\cal B}(B_s\to \mu^- \mu^+)=3.78^{+0.15}_{-0.10} \times 10^{-9}$~\cite{Buras:2022qip}, slightly higher than the current experimental measurement ${\cal B}(B_s\to \mu^- \mu^+)=(3.01\pm 0.35) \times 10^{-9}$~\cite{ParticleDataGroup:2022pth}. The introduction of right-handed quark currents helps alleviate this tension.

Because the right-handed quarks and leptons in the SM and the introduced dark quark are $SU(2)_L$ singlet, we can employ an $SU(2)_L$ singlet electrically charged dark scalar to couple these particles. As a result, the right-handed currents for the $b\to s$ transition can be generated from one-loop Feynman diagrams. Moreover, through mixing with the inert Higgs doublet, the left- and right-handed leptons can couple to the physical inert charged scalars. This results in the corrections to lepton $g-2$ being linear in the lepton mass. Therefore, the muon $g-2$ can be significantly enhanced in the model. Based on the above analysis, the additional dark particles introduced in the model to explain the neutrino measurements and the muon $g-2$, to fit ${\cal B}(B_s \to \mu^- \mu^+)$, and to enhance the $d_i \to d_j \nu \bar\nu$ processes are: the non-leptonic inert Higgs doublet, the singlet vector-like up-type dark quark, and the singlet dark-charged scalar.

The paper is structured as follows.  We set up the scotogenic model and derive the Yukawa and relevant gauge couplings in Sec.~\ref{sec:model}.  Utilizing the obtained Yukawa couplings and scalar mixings, we formulate the loop-induced neutrino mass matrix and lepton $g-2$ in Sec.~\ref{sec:processes}. The constraints from $b\to s \gamma$ and $|\Delta F| = 2$ arising from box diagrams are analyzed. In addition, the effective Hamiltonian for $d_i \to d_j (\nu \bar\nu, \mu^- \mu^+)$, which arises from the $Z$-penguin diagrams, is derived in this section. Based on the new interactions, the BRs for $B\to K^{(*)} \nu \bar\nu$, $K\to \pi \nu \bar\nu$, $B\to X_s \ell^- \ell^+$, and $B_s\to \mu^- \mu^+$ are computed in Sec.~\ref{sec:obs}. The detailed numerical analysis and discussions of the phenomenological results are shown in Sec.~\ref{sec:num}. The findings of this study are summarized in Sec.~\ref{sec:sum}.

\section{The model and couplings} \label{sec:model}

As stated in the introduction, the Majorana neutrino mass can be radiatively generated in the scotogenic model if lepton number violation originates from the coupling of the leptonic inert Higgs doublet to the non-leptonic inert Higgs and the SM Higgs doublets in the scalar potential. Moreover, by introducing an $SU(2)_L$ singlet vector-like up-type quark, we can have interesting phenomenological contributions to the FCNC $B$ and $K$ decays. Therefore, to investigate the impacts of new physics on the processes of $d_i\to d_j  (\nu \bar\nu, \ell^- \ell^+)$ while avoiding the strict constraints from the $B\to X_s \ell^- \ell^+$ and $B_s \to \mu^+ \mu^-$ decays, we extend the SM by adding two Higgs doublet $\eta_{1,2}$, one charged scalar $\chi^\pm$, and singlet vector-like neutral leptons $N_{L,R}$ and up-type quarks $T_{L,R}$, one for each chirality.

It is found that with appropriate charge assignment to the new particles, a global $U(1)_X$ dark symmetry exists in the model. Since a $Z'$-guage boson is not necessary for the study, we do not gauge the $U(1)_X$ symmetry. For clarity, we show the representations and assignments of $U(1)_X$ charge and lepton number as follows:
\begin{align}
\begin{split}
\eta_{1} & \sim  (2, \, 1,\, q_X,\, 0)\,,~ \eta_{2}\sim (2,\,1,\, -q_X, -2)~, \\
N_{L,R} & \sim (1,\,0,\, q_X,\, 1)\,,~T_{L,R}\sim (1,\, 4/3,\, q_X,\, 0)\,,~ \chi^+\sim (1,\, 2,\, q_X,\, 0)\,, 
 \label{eq:rep}
\end{split}
\end{align}
where the numbers in the parentheses denote in sequence the $SU(2)_L$ representation, the $U(1)_Y$ charge, the $U(1)_X$ charge, and the lepton number. We note that both $\eta_2$ and $N$ carry the lepton number. Moreover, to ensure the stability of a DM candidate in the model, we assume that the $U(1)_X$ is exact, and the $\eta_{1,2}$ are assumed to have no nonzero vacuum expectation values (VEVs). Hence, the masses of the dark-charged scalars do not originate from the electroweak symmetry breaking.

We will focus on the loop-induced quark flavor-changing processes in this work.  Details of the scalar potential, the scalar mixings, and the scalar mass spectra are presented in Appendix~\ref{app:potential}.  The main free parameters associated with the scalar sector are the masses of the inert charged Higgs bosons.

\subsection{Yukawa couplings}

Based on the representations and charge assignments in Eq.~(\ref{eq:rep}), the Yukawa couplings for the new particles are given by:
\begin{equation}
\begin{split}
 -{\cal L}_Y = & 
 \overline L \, {\bf y}_1 \, \tilde\eta_1 N_{ R} + \overline L\, {\bf y}_2\, \tilde\eta_2 N^C_{ L }  + \overline{\ell_R}\, {\bf y}_{\ell}\, N_L \chi^- + m_{N} \overline N_{L} N_{R}  \\
 &  + \overline{Q_L}\, {\bf y}^R_T \, \tilde\eta_1 T_R  + \overline{d_R} \, {\bf y}^L_{T} \, T_L \chi^- + m_T \overline{ T_L} T_R+  \mbox{H.c.}~, 
 \label{eq:yukawa}
 \end{split}
 \end{equation}
where the flavor indices are suppressed, $L$ and $Q_L$ denote respectively the left-handed lepton and quark doublets in the SM, ${\tilde\eta_j} = i \tau_2 \eta^*_j$, $N^C= C  \bar N^T$ is the charge conjugation of $N$, and $m_N~(m_T)$ is the mass of $N~(T)$.  After electroweak symmetry breaking, we introduce the unitary flavor-mixing matrices $V^\ell_{R,L}$ and $V^q_{R,L}$ to diagonalize the charged lepton and quark mass matrices. Since the neutrinos are still massless at the tree level, we can absorb the lepton flavor-mixing matrices to the Yukawa couplings ${\bf y}_{1,2, \ell}$.  If we rotate away the weak phase of ${\bf y}_{1}$, both ${\bf y}_{2}$ and ${\bf y}_{\ell}$ in general contain complex parameters.  Thus, ${\bf y}_2$ can lead to a complex Majorana neutrino mass matrix through radiative corrections. Using the physical charged and neutral scalar states defined in Eqs.~(\ref{eq:mixing}) and (\ref{eq:omatrix}),  the lepton Yukawa couplings can be obtained as:
\begin{align}
\begin{split}
-{\cal L}^\ell_Y =& \overline{\nu_L} \, {\bf y}_1\,  N_R\,  \left( c_\phi (S_1-i A_1) -s_\phi (S_2 + iA_2)\right)  \\
 & +  \overline{\nu_L} \, {\bf y}_2\,  N^C_L\,  \left( s_\phi (S_1+i A_1) +c_\phi (S_2 - iA_2)\right) -  \overline{\ell_L} \, {\bf y}_{2}\, N^C_L \, \eta^-_2  \\
& -  \overline{\ell_L} \, {\bf y}_{1} \, N_R \left(c_\theta H^-_1 - s_\theta H^-_2 \right) +  \overline{\ell_R}\,  {\bf y}_{\ell} \,  N_L\left(s_\theta H^-_1 + c_\theta H^-_2 \right) + m_N \overline{N_L} N_R + \mbox{H.c.} \,,  
\label{eq:yuka_lepton}
\end{split}
\end{align}
where $\theta$ denotes the mixing angle between $\eta^\pm_1$ and $\chi^\pm$.
Besides the masses of neutral scalars, we will show later that the loop-induced Majorana neutrino mass matrix depends on ${\bf y}_{1,2}$ and the angle $\theta$. 

Before electroweak symmetry breaking, the weak phases of ${\bf y}^{R,T}_T$ can be rotated away by redefining the phases of the quark fields $u_L$ and $d_{L,R}$.  After the symmetry breaking, when the quark-flavor mixing matrices are introduced for diagonalizing the quark mass matrices, we can redefine $V^u_L {\bf y}^R_T$ as ${\bf y}^R_T$ in the $\bar u_L {\bf y}^R_T T_R \eta^{0*}_1$ term. As a result, the term related to the left-handed down-type quark becomes $\bar d_L V^\dag {\bf y}^R_T T_R \eta^-_1$, where $V=V^u_L V^{d\dag}_L$ represents the Cabibbo-Kobayashi-Maskawa (CKM) matrix. In terms of physical quark and scalar states, the Yukawa couplings of the new quark in the model are expressed as:
\begin{align}
\begin{split}
-{\cal L}^q_Y   =  & \overline{u_L} \, {\bf y}^R_T \, T_R \, 
\left[ c_\phi (S_1-i A_1) -s_\phi (S_2 + iA_2) \right]  
\\
 & + \bar d \, \left( {\bf C}^{R k}_{Ti} P_R +  {\bf C}^{L k}_{T} P_L \right) T H^-_k + m_T \overline{T_L} T_R + \mbox{H.c.} 
 \label{eq:yuka_quark}
\end{split}
\end{align}
Here, ${\bf C}^{R1(2)}_T = -c_\theta(s_\theta)  {\bf Y}^R_T$, and ${\bf C}^{L1(2)}_T = s_\theta (c_\theta) {\bf Y}^L_T$, with 
 \begin{equation}
  {\bf Y}^R_T=  V^\dag {\bf y}^R_T\,, ~  {\bf Y}^L_T= V^{d}_R \, {\bf y}^L_T\,. \label{eq:YLR}
 \end{equation}
From Eq.~(\ref{eq:yuka_quark}), it can be seen that the down-type quarks only couple to the charged Higgses.  Although the Yukawa couplings ${\bf y}^{R,L}_T$ can be tightly restrained by the up-type quark processes mediated by the scalars $S_i$ and pseudoscalars $A_i$, such as $D-\bar D$ mixing, the constraint is essentially close to that from the $K-\bar K$ mixing. Without loss of generality, we will focus on the charged Higgs-mediated phenomena.

If the up-type quarks in the weak eigenstates are initially aligned with their physical states (i.e., $V^u_L=\mathbbm{1}$), we then have $V=V^{d\dag}_L$, which is well-determined in experiments. Since there is no information on the right-handed quark flavor mixings, $V^d_R$ is essentially an unspecified unitary matrix. To reduce the number of free parameters, we adopt the assumption that $V^d_R = V^d_L$. Indeed, the assumption can be realized in the left-right symmetric model or model with a Hermitian quark mass matrix. The unique CP-violating phase in the quark sector then arises from the Kobayashi-Maskawa (KM) phase in the CKM matrix. We will apply the assumption in the numerical analysis.

\subsection{Gauge couplings}

In addition to the Yukawa couplings, the alternative interactions essential for the study of the FCNC $B$ and $K$ decays are the gauge couplings of the photon ($A_\mu$) and $Z$-boson to $T$, $\chi^\pm$, and $\eta^+_1$. Being an $SU(2)_L$ singlet and charged under $U(1)_X$, $T$ does not mix with the SM quarks or couple with the $W$ boson.

To obtain the relevant guage interactions, we write the covariant derivatives of $T$, $\chi^\pm$ and $\eta_i$ as:
\begin{align}
\begin{split}
  D_\mu T &= (\partial_\mu + i Q_T g'  B_\mu) T
   ~, \\
  D_\mu \chi^+ &= (\partial_\mu + i  g'  B_\mu) \chi^+  
  ~, \\
  D_\mu \eta_1 & = \left(\partial_\mu +  i \frac{g}{2} \vec{\tau} \cdot \vec W_\mu + i \frac{g'}{2} B_\mu \right) \eta_1
  ~,
\end{split}
\end{align}
where the hypercharges of $Y_T=4/3=2 Q_T$, $Y_{\chi^+}=2$ and $Y_{\eta_1}=1$ have been explicitly used. Because $\eta_2$ does not couple to quarks, we refrain from showing its gauge couplings. Using the weak mixing angle, defined by $\cos\theta_W=g/\sqrt{g^2+g'^2}$ and $\sin\theta_W=g'/\sqrt{g^2+g'^2}$, and the relation of $g\sin\theta_W=g'\cos\theta_W=e$, we parametrize the photon and $Z$-boson states as:
\begin{align}
\begin{split}
 A_\mu &= c_W B_\mu + s_W W^3_\mu~,
  \\
 Z_\mu &= -s_W B_\mu + c_W W^3_\mu~,
\end{split}
\end{align}
with $c_W \equiv \cos\theta_W$ and $s_W \equiv \sin\theta_W$. The gauge couplings of $A_\mu$ and $Z_\mu$ to the heavy new quark can be found as:
 \begin{equation}
 {\cal L}_{TTV} = - e Q_T \bar T \gamma_\mu T A^\mu - \frac{g}{2 c_W} \left(-2Q_T s^2_W \right) \bar T \gamma_\mu T Z^\mu~. 
 \end{equation}
Note that as $T$ is a vector-like quark, it has a vectorial coupling to the $Z$ gauge boson.

For the gauge couplings of the charged scalars $H^\pm_i$, they can be obtained as:
 \begin{align}
  {\cal L}_{H^-H^+V}  
  = &~ 
  i e A^\mu \sum^2_{i=1} (H^+_i \partial_\mu H^-_i  - H^-_i \partial_\mu H^+_i) \nonumber \\
  &+   i \frac{g}{2c_W} c^Z_{ij} Z^\mu   \left( (\partial_\mu H^-_i)  H^+_j - H^-_i \partial_\mu H^+_j \right)  \,, \label{eq:scalar_gauge}
  \end{align}
where  $c^Z_{11}=c^2_{\theta}-2 s^2_W$, $c^Z_{12}=c^Z_{21}=c_\theta s_\theta$, and $c^Z_{22}=s^2_\theta-2 s^2_W$. 
Because $\eta^\pm_1$ and $\chi^\pm $ mix together and belong to different $SU(2)_L$ representations, the $Z$ couplings to these charged Higgs fields are not diagonal. We note that the gauge couplings of scalars $S_i$ and pseudoscalars $A_i$ to the $Z$ gauge boson can be expressed as ${\cal L}_{\rm kin} \supset g/(2 c_W) \left(A_i \partial_\mu S_i - S_i \partial_\mu A_i \right) Z^\mu$. From the results shown in Appendix~\ref{app:potential}, $S_i$ and $A_i$ are degenerate due to the $U(1)_X$ symmetry. Therefore, these bosons cannot be the DM candidates; otherwise, the scalar boson scattering off the nucleon, $S_i {\cal N} \to A_i {\cal N}$ or its inverse process, mediated by the $Z$ gauge boson will lead to too large a cross section that has already been excluded by the DM direct detection. Thus, the neutral Dirac fermion $N$ is the DM candidate in this model. Utilizing the Yukawa couplings in Eq.~(\ref{eq:yuka_lepton}), the annihilation cross section of $N\bar N\to f_{\rm SM} \bar f_{\rm SM}$ can accommodate the observed DM relic density when $m_N \lesssim 600$~GeV, as detailed in Ref.~\cite{Chen:2022gmk}.

\section{Loop-induced  processes} \label{sec:processes}

In the following, we examine various loop-mediated processes that receive additional contributions from the new particles in the model.

\subsection{Radiative neutrino mass and muon $g-2$} \label{subsec:gminus2}

According to the lepton Yukawa couplings shown in Eq.~(\ref{eq:yuka_lepton}), the Majorana neutrino mass matrix elements mediated by the neutral scalars and $N$ can be obtained as:
\begin{equation}
m^\nu_{ij}  = \frac{\sin(2\phi)}{32\pi^2}  \overline y_{ij} m_{N} \left[ \frac{m^2_{S_1}}{m^2_{S_1} -m^2_{N}} \ln\left( \frac{m^2_{S_1}}{m^2_{N} }\right) - \frac{m^2_{S_2}}{m^2_{S_2} -m^2_{N}} \ln\left( \frac{m^2_{S_2}}{m^2_{N} }\right) \right]~,  \label{eq:nu_mass}
\end{equation}
where we have included the pseudoscalar $A_i$ contributions, used the mass relation $m_{A_i}=m_{S_i}$, and defined the symmetric Yukawa couplings $\overline y_{ij}$  in flavor indices as $\overline{y}_{ij} =  y^{*}_{1i} y^{*}_{2j} + y^{*}_{2i} y^{*}_{1j}$. For illustration purposes, we take $m_{S_1}=600$~GeV, $m_{S_2}=800$~GeV, $m_N=300$~GeV, $\phi\sim {\cal O}(10^{-7})$, and ${\bf y}_{1,2}\sim {\cal O}(10^{-2})$ which is the order of $\tau$ lepton Yukawa coupling in the SM, and obtain $m^\nu_{ij} \sim {\cal O}(10^{-2})$~eV.  To have more impacts on the lepton flavor-violating processes, such $\mu\to e \gamma$, $\mu\to 3 e$, and $\mu-e$ conversion, one can take ${\bf y}_{1,2}$ of ${\cal O}(1)$ while keeping $\phi\sim 10^{-11}$ or $\lambda_5 \sim {\cal O}(10^{-10})$. In the limit of $\lambda_5=0$, the lepton number symmetry is restored. Therefore, a small $\lambda_5$ can be regarded as technically natural~\cite{tHooft:1979rat}.

When considering the scheme with ${\bf y}_1\sim {\cal O}(1)$, one might anticipate significant effects on $d_i \to s (\nu \bar\nu, \ell^- \ell^+)$ from box diagrams, where the $T$-quark, $N$-lepton, and charged Higgses run inside the loops. However, unlike the $Z$-penguin diagrams that induce dimension-4 $d_i$-$s$-$ Z^*$ couplings, the effective operators arising from the box diagrams for $d_i \to s (\nu \bar\nu, \ell^- \ell^+)$ are of dimension-6. In other words, the effective Wilson coefficients resulting from ${\bf y}_{1}$ are suppressed by $m^2_W/m^2_T$. Consequently, the lepton Yukawa couplings cannot have a significant effect on the quark flavor-changing processes.

Using the Yukawa couplings in Eq.~(\ref{eq:yuka_lepton}), the radiative corrections to the muon magnetic dipole moment mediated by $H^+_{1,2}$ and $N$ can be obtained as:
\begin{equation}
\Delta a_\mu =  -\frac{m_\mu s_{2\theta}}{16 \pi^2 } m_N \operatorname{Re}(y_{\ell 2} y^*_{12}) \left( \frac{J_1(w_{H^+_1}) }{m^2_{H^+_1}}- \frac{J_1(w_{H^+_2})}{m^2_{H^+_2}}\right)\,,
\end{equation}
where $w_{H^+_i} \equiv m^2_N /m^2_{H^+_i}$, and $J_1$ is a loop integral, defined as:
   \begin{equation}
   J_1(w)  =  \frac{1+w }{2(1-w)^2 } + \frac{w\ln w}{(1-w)^3} \,. \label{eq:J1}
   \end{equation}
Because $\eta^+_1$ and $\chi^+$ couple to the left-handed and right-handed leptons, respectively, the resulting $\Delta a_\mu$ is proportional to $m_\mu$. The mass insertion factor occurring in the $N$ propagator further enhances $\Delta a_\mu$. In addition, without introducing $\chi^+$, the contribution to $\Delta a_\mu$ from $\eta^+_1$ would always have a negative sign. Assuming $m_{H^+_1}=600$~GeV, $m_{\chi^+_1}=800$~GeV, $m_N=300$~GeV, $s_\theta=0.1$, and $\operatorname{Re}(y_{\ell 2} y^*_{12})=0.2$ as an illustration, we obtain $\Delta a_\mu \simeq  2.2\times 10^{-9}$.

The above numerical estimate demonstrates that the model can readily accommodate the observed neutrino mass and the muon $g-2$ anomaly. Since the purely lepton-related processes in the model are similar to the study in Ref.~\cite{Chen:2022gmk}, a detailed analysis can be found therein. This study primarily focuses on exploring rare quark flavor-changing processes.

\subsection{$b\to s \gamma$}

Due to the precision measurement of $b\to s\gamma$ decays, the new physics effects contributing to $b\to s (\nu \bar \nu, \, \ell^- \ell^+)$ are severely constrained. In this subsection, we examine the influence of new couplings on the $b\to s \gamma$ decays.

The effective Hamiltonian for $b\to s \gamma^{(*)}$ from photon-penguin diagrams mediated by $H^\pm_{1,2}$ and $T$ can be parametrized as:
\begin{equation}
{\cal H}^{NP}_{b\to s\gamma} = k^2 \, \bar{s}\, \gamma^\mu \, ( A_L P_L + A_R P_R) \, b\, A_\mu + \bar{s}\, \sigma^{\mu\nu}  \, (B_L P_L + B_R P_R)\, b\, F_{\mu\nu}\,, \label{eq:btosga}
\end{equation}
where $k_\mu$ is the momentum of the emitted photon, and $F_{\mu\nu}$ is the field strength tensor of the electromagnetic field. For an on-shell photon, i.e., $k^2=0$, only the dipole operators contribute. By dimensional analysis, since $A_L$ and $A_R$ are associated with vector currents and the chiralities in the initial and final states of quarks are the same, $A_L$ and $A_R$ are proportional to $1/m^2_T$ if $T$-quark is the heaviest particle in the model. As a result, the processes $b\to s\, \ell^-\, \ell^+$  from the off-shell photon are indeed suppressed and negligible.

The situation for the dipole operators is more complicated. By dimensional analysis, $B_{L,R}$ can depend on $m_b/m^2_T$ and $1/m_T$, where $m_b$ arises from the application of the equation of motion or chirality flip in the $b$-quark propagator. Since $m_s\ll m_b$, we take $m_s\approx 0$. In terms of the weak states of $\eta^+_1$ and $\chi^+$, one can easily understand that the contribution from each of $\eta^+_1$ and $\chi^+$ results in the dependence of $m_b/m^2_T$. Since $\eta^+_1$ and $\chi^+$ only couple to the left-handed and the right-handed quarks, respectively, to flip the chirality of the $b$-quark in the tensor-type weak currents, the mass effect of $b$-quark must appear.  Hence, the dimension-4 Hamiltonian mediated by $\eta^+_1/\chi^+$ leads to $B_{R,L} \propto m_b/m^2_T$.

When considering the mixing effect of $\eta^+_1$ and $\chi^+$, the chirality flip in the tensor-type current is automatically satisfied.  Moreover, instead of the $m_b$ effect, a mass insertion factor appears in the propagator of $T$ that runs in the photon-penguin loop.  Consequently, the contributions from the $\eta^+_1$-$\chi^+$ mixing are expected to be $B_{L,R}\propto c_\theta s_\theta/m_T$, which are larger than those from the individual $\eta^+_1$ and $\chi^+$ contributions. The dominant effects of $B_{L,R}$ can then be expressed as:
  \begin{align}
  \begin{split}
  B_{L}  & = \frac{(2+Q_T) c_\theta s_\theta}{32 \pi^2 m_T} Y^{L}_{T2} Y^{R*}_{T3}  \left( J_1(z_{H^+_2}) - J(z_{H^+_1})\right) 
  \,, \\
   B_{R}  & = \frac{(2+Q_T) c_\theta s_\theta}{32 \pi^2 m_T} Y^{R}_{T2} Y^{L*}_{T3} \left( J_1(z_{H^+_2}) - J(z_{H^+_1})\right)  
   \,, \label{eq:B_LR}
  \end{split}
  \end{align}
with $z_{H^+_i} \equiv m^2_{H^+_i} /m^2_T$. To avoid the constraint on ${\bf Y}^{R,L}_T$ from $b\to s \gamma$, we can either consider $s_\theta$ to be sufficiently small or $m_{H^+_1} \sim m_{H^+_2}$ for a subtle cancellation in $J_1(z_{H^+_2}) - J(z_{H^+_1})$. For numerical illustration purposes,  we rewrite Eq.~(\ref{eq:btosga}) in terms of the standard magnetic dipole operators as:
\begin{align}
{\cal H}^{\rm NP}_{b\to s \gamma} & = -\frac{G_F V^*_{ts} V_{tb}}{\sqrt{2}} \left(C^{\rm NP}_7 O_{7\gamma} + C^{\prime \rm NP}_{7\gamma} O^{\prime}_{7\gamma}\right)\,, \\
O^{(\prime)}_{7\gamma} & =  \frac{e m_b}{ 4\pi^2} \bar s\, \sigma_{\mu \nu} P_{R(L)} \, b\, F^{\mu \nu}\,. \nonumber 
   \end{align}
The SM results are $C^{\rm SM}_{7\gamma}\approx -0.3$ and $C^{\prime \rm SM}_{7\gamma}\approx 0$.  Using the typical values of the parameters: $m_T=1.2$~TeV, $m_{H^+_1}=0.6$~TeV, $m_{H^+_2}=0.8$~TeV, $s_\theta=0.1$, and $Y^{L}_{T2} Y^{R*}_{T3}\sim Y^{L}_{T2} Y^{R*}_{T3}\sim 0.1$, we obtain $|C^{(\prime)\rm NP}_{7\gamma}/C^{SM}_{7\gamma}|\sim  0.19$. As no anomalous signals are found in the $b\to s \gamma$ processes, we can simply suppress $C^{(\prime)\rm NP}_{7\gamma}$ without further limiting the Yukawa couplings ${\bf Y}^{R,L}_{T}$, which are the primary factors contributing to the $|\Delta F| = 1$ and $|\Delta F| = 2$ ($F=K$, $B_d$, and $B_s$) processes in the model.

\subsection{$|\Delta F| = 2$ from box diagrams}

Among loop-induced FCNCs involving the down-type quarks,  the essential and most well-measured processes are $K$ and $B_q$ ($q=d,s)$ meson oscillations, characterized by the mass differences between their mass eigenstate, denoted by $\Delta m_{K,\, B_q}$. To evaluate the impact of new physics effects in the model on the $|\Delta F| = 2$ processes, we derive the $\Delta m_F$, including the SM contributions, as follows.

\begin{figure}[phtb]
\begin{center}
\includegraphics[scale=1]{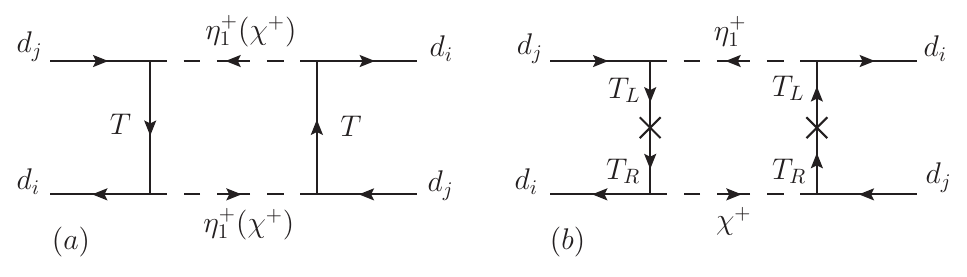}
\caption{ Selected Feynman diagrams for $|\Delta F| = 2$ processes. }
\label{fig:boxes}
\end{center}
\end{figure}

Since there is no FCNC coupling at the tree level, the $|\Delta F| = 2$ processes have to arise from the box diagrams, and the representative Feynman diagrams are sketched in Fig.~{\ref{fig:boxes}, where the left-handed plot shows the mediation of $\eta^+_1$-$\eta^+_1$ and $\chi^+$-$\chi^+$, and the right-handed plot shows that of $\eta^+_1$-$\chi^+$.  Since the four fermions in the external legs of the box diagrams lead to dimension-6 operators, the associated effective coefficients in the induced Hamiltonian have to be proportional to $1/m^2_T$. Although a $m^2_T$ factor appears at the propagators of $T$ via the mass insertions, the dependence of $1/m^2_T$ is not changed in Fig.~\ref{fig:boxes}(b) because the denominator of the loop integrand has an eighth power of momentum in four-dimensional momentum integral. Therefore, the final result still shows the dependence of $1/m^2_T$.  
Using the Yukawa couplings shown in Eq.~(\ref{eq:yuka_quark}), the effective Hamiltonian for $|\Delta F| = 2$ can be obtained as:
\begin{align}
\begin{split}
{\cal H}^{\rm NP}(|\Delta F| = 2) = & \frac{c^4_\theta}{4 (4 \pi)^2 m^2_T}   (Y^R_{Ti} Y^{R*}_{Tj} )^2 J_1 (z_{H^+_1}) \, \overline{d_i} \gamma_\mu P_L d_j \, \overline{d_i} \gamma^\mu P_L d_j \\
& +  \frac{c^4_\theta}{4 (4 \pi)^2 m^2_T}  
(Y^L_{Ti} Y^{L*}_{Tj})^2   J_{1}(z_{H^+_2}) \, \overline{d_i} \gamma_\mu P_R d_j \, \overline{d_i} \gamma^\mu P_R d_j  \\
& - \frac{c^4_\theta}{(4 \pi)^2 m^2_T} \left( Y^L_{Ti} Y^{L*}_{Tj} Y^{R}_{Ti} Y^{R*}_{Tj} \right) J_{2} (z_{H^+_1}, z_{H^+_2}) \, \overline{d_i} P_L d_j \, \overline{d_i} P_R d_j \,, \label{eq:HDF=2}
\end{split}
\end{align}
where all involved Feynman diagrams have been considered, the small contributions from the dependence of $s_\theta$ and $m_b$ are neglected, and the flavor index pairs $(i ,\, j)=(2,\, 1)$, $(3,\, 1)$, and $(3,\, 2)$  correspond to the $K$, $B_d$, and $B_s$ mesons, respectively.  The loop integral function $J_1$ is the same as that in Eq.~(\ref{eq:J1}),  and $J_2$ is given by
\begin{equation}
J_2(w,x) = - \frac{1}{(1-w)(1-x)} - \frac{1}{w-x} \left[ \frac{w \ln w}{(1-w)^2} - \frac{x \ln x}{(1-x)^2}\right]\,.
\end{equation}

The mass differences between the heavy and light neutral $K/B_q$ mesons are defined as:
 \begin{align}
 \begin{split}
 \Delta m_K  & = 2\, \operatorname{Re}(M^K_{12})
 = 2\,  \operatorname{Re}(\langle \bar K | {\cal H}(|\Delta S| = 2) | K\rangle)
 \,, \\
 \Delta m_{B_q} & = 2\,  |M^{B_q}_{12}| 
 = 2\,  | \langle \bar B_q | {\cal H}(|\Delta S| = 2) | B_q \rangle|
 \,. \label{eq:DmF}
 \end{split}
 \end{align} 
To estimate the matrix elements of $M^{K}_{12}$ and $M^{B_q}_{12}$ contributed from the new physics effects, we apply the results obtained in Ref.~\cite{Buras:2001ra}, where the QCD renormalization group effects from a high energy scale to a low $\mu$ scale are included. To apply the results from Ref.~\cite{Buras:2001ra}, the $|\Delta F| = 2$ Hamiltonian is parametrized by
 \begin{equation}
 {\cal H}(|\Delta F| = 2) = \frac{G^2_F}{16 \pi^2} m^2_W \sum_i C_i (\mu) Q_i \,, 
 \end{equation}
where the operators $Q_i$ are defined as~\cite{Buras:2001ra}:
 \begin{subequations} \label{eq:4ops}
 \begin{align}
 Q^{\rm VLL}_1 & = (\overline{d_i} \gamma_\mu P_L d_j)  (\overline{d_i} \gamma^\mu P_L d_j)\,, \\ 
 Q^{\rm LR}_1 & = (\overline{d_i}  \gamma_\mu P_L d_j)  (\overline{d_i}  \gamma^\mu P_R d_j) \,, \\ 
 Q^{\rm LR}_2 & = (\overline{d_i}  P_L d_j)  (\overline{d_i}  P_R d_j) \,, \\ 
 Q^{\rm SLL}_{1} & = (\overline{d_i}  P_L d_j)  (\overline{d_i}  P_L d_j) \,, \\ 
 Q^{\rm SLL}_2 & = (\overline{d_i}  \sigma_{\mu\nu} P_L d_j)  (\overline{d_i}  \sigma^{\mu \nu} P_L d_j) \,. 
 \end{align}
\end{subequations}
Here we have suppressed the color indices.  Since the matrix elements of the operators VRR and SRR are the same as those of VLL and SLL, we do not explicitly show these operators in Eq.~(\ref{eq:4ops}). The master formula for the meson-antimeson matrix element is expressed as~\cite{Buras:2001ra}:
 \begin{align}
 \begin{split}
M^{\rm F, NP}_{12}  = & \frac{G^2_F}{16 \pi^2} m^2_W \frac{F^2_F m_F}{3} \left[ P^{\rm VLL}_1 \left( C^{\rm VLL}_1 + C^{\rm RLL}_{1} \right)+ P^{\rm LR}_1 C^{\rm LR}_1 + P^{\rm LR}_{2} C^{LR}_{2} \right.  \\
 &  \left. +  P^{\rm SLL}_1 \left( C^{\rm SLL}_{1} + C^{\rm SRR}_1\right) + P^{\rm SLL}_{2} \left( C^{\rm SLL}_2 + C^{\rm SRR}_2 \right)\right]\,, \label{eq:Master}
 \end{split}
 \end{align}
where $P^\alpha_i$ include the non-perturbative bag parameters and QCD running factors,  $C^\alpha_i$ denote the effective coefficients at the high energy scale, and $F_F$ is the decay constant of meson $F$. The values of $P^\alpha_i$ are shown in Table~\ref{tab:P-values}~\cite{Buras:2001ra}.  We note that with the exception of $P^{\rm VLL}_1$, the values of $P^\alpha_i$ in the $K$ meson are one order of magnitude larger than those in the $B_q$ meson due to the factor of $m^2_F/(m_{d_i}+ m_{d_j})^2$. This factor in $K$ is approximately 20 times larger than that in $B_q$.

\begin{table}[thp]
 \caption{ Values of $P^\alpha_i$ used to estimate the matrix elements of the $K$, $B_d$, and $B_s$ mesons. }
\begin{center}
\begin{tabular}{c|ccccc} \hline \hline
& ~~$P^{\rm VLL}_1$~~ & ~~$P^{LR}_1$~~  &  ~~$P^{\rm LR}_2$~~ & ~~$P^{SLL}_1$~~ &~~ $P^{SLL}_2$~~ \\ \hline
~~$K$ ~~& 0.48 & $-36.1$ & $59.2$ & $-18.1$  & $-32.2$ \\ \hline
 $B_d$ & 0.84 & $-1.62$ & $2.46$ & $-1.47$ & $-2.98$\\ \hline
 $B_s$ &  0.94 & $-1.83$ & $2.78$ & $-1.66$ & $-3.37$\\ \hline \hline  
\end{tabular}
\end{center}
\label{tab:P-values}
\end{table}

From Eq.~(\ref{eq:HDF=2}), the involved four-fermion operators are $Q^{\rm VLL, VRR}_1$ and $Q^{LR}_2$. Using the master formula in Eq.~(\ref{eq:Master}), the resulting mixing matrix element for $F$ transition to $\bar F$ can then be written as:
 \begin{align}
 \begin{split}
 M^{\rm F, NP}_{12}  = & \frac{F^2_F m_F c^4_\theta}{48 \pi^2 m^2_T}  
 \Bigg[ \frac{P^{VLL}_1}{4} \left( (Y^R_{Ti} T^{R*}_{Tj})^2 J_1(z_{H^+_1}) +  (Y^L_{Ti} T^{L*}_{Tj})^2 J_1(z_{H^+_2})\right)  \\
 &  - \,  P^{LR}_2 \left( Y^{R}_{Ti} Y^{R*}_{Tj} \right)  \left( Y^L_{Ti} Y^{L*}_{Tj}\right) J_{2} (z_{H^+_1}, z_{H^+_2})  \Bigg]\,.
 \end{split}
 \end{align}
In order to combine the SM contributions, we write the SM results as~\cite{Buchalla:1995vs}:
 \begin{align}
 \begin{split}
 M^{\rm K, SM}_{12} & = \frac{G^2_F m^2_W}{12 \pi^2} F^2_K m_K \hat{B}_K \left[ \lambda^2_c \xi_1 S_0(x_c)+ \lambda^2_t \xi_2 S_0 (x_t) +2 \lambda_c \lambda_t \xi_3 S_0(x_c, \, x_t)\right] 
 \\
 M^{\rm B_q, SM} _{12} & = \frac{G^2_F m^2_W}{12 \pi^2}  \lambda^2_t F^2_{B_q} m_{B_q} \hat{B}_{B_q} \xi_B S_0 (x_t) \,,  \label{eq:MDM12}
 \end{split}
 \end{align}
 where $\lambda_k = V^*_{k i} V_{k j}$; $\xi_{1, 2, 3}=(1.87\pm 0.76,\, 0.5765\pm 0.0065,\, 0.496 \pm  0.047)$~\cite{Brod:2019rzc} and $\xi_B=0.5510\pm 0.0022$~\cite{Buras:1990fn,Lenz:2010gu} are the 
 QCD correction factors, and  $\hat B_{F}$ is the renormalization scale-independent bag parameter. With $x_f \equiv m^2_f/m^2_W$, the $S_0$ functions are read as~\cite{Buchalla:1995vs}:
 \begin{align}
 \begin{split}
 S_0(x_c) & = x_c\,, \\
 S_0(x_t) &=  \frac{4 x_t -11 x^2_t + x^3_t}{4 (1-x_t)^2} - \frac{3 x^3_t \ln x_t}{2 (1-x_t)^3}\,, \\
 S_0(x_c, x_t) &= x_c \left( \ln\frac{x_t}{x_c} - \frac{3 x_t}{4(1-x_t)} - \frac{3 x^2_t \ln x_t}{4 (1-x_t)^2}\right)\,.
 \end{split}
 \end{align}
Accordingly, the transition matrix element by combining the SM result and the charged-Higgs mediated effects is given by $M^{F}_{12}=M^{\rm F, SM}_{12} + M^{\rm F, NP}_{12}$. Using Eq.~(\ref{eq:DmF}), we can obtain $\Delta m_{F}$.

\subsection{$Z$-penguin induced  $d_i \to d_j  ( \nu \bar\nu, \ell^- \ell^+$)}  

As alluded to earlier, the $d_i \to d_j \ell^- \ell^+$ processes mediated through the photon-penguin diagrams have negligible contributions due to the $1/m^2_T$ suppression. Box diagrams, mediated by $T$ and $N$ and governed by the Yukawa couplings ${\bf y}_{1}$, can also induce $d_i \to d_j \ell^- \ell^+$. However, similar to the $|\Delta F| = 2$ case, the induced dimension-6 four-fermion operators of $d_i d_j \ell^- \ell^+$ are suppressed by $1/m^2_T$. Therefore, the primary contributions to $d_i \to d_j \ell' \overline{\ell'}$ ($\ell'=\nu_\ell, \ell^\pm$) in the model are predominantly from the $Z$-penguin diagrams. The dominant Feynman diagram is depicted in Fig.~\ref{fig:Z_penguin}.

\begin{figure}[phtb]
\begin{center}
\includegraphics[scale=1]{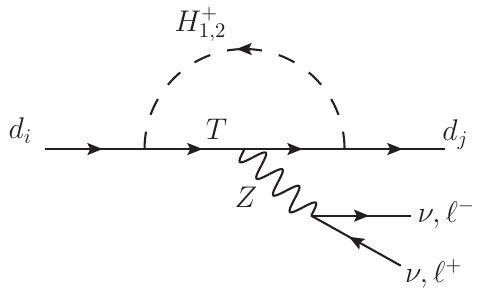}
 \caption{ Dominant $Z$-penguin Feynman diagrams for $d_i\to d_j (\nu \bar \nu,\, \ell^- \ell^+)$. }
\label{fig:Z_penguin}
\end{center}
\end{figure}

Analogous to Eq.~(\ref{eq:btosga}), the loop-induced $d_i d_j Z^*$ vertices generally contain structures of vector and tensor currents.  Through dimensional analysis and consideration of quark chirality, it can be found that the effective coefficients associated with the tensor currents are proportional to $c_\theta\, m_{d_i}/m^2_T$ and $s_\theta  /m_T$, where the former corresponds to the contributions from $\eta^+_1(\chi^+)$, and the latter is from the mixing of $\eta^+_1$ and $\chi^+$. As the tensor operator $\overline{d_j} \sigma_{\mu\nu} d_i Z^{\mu \nu}$ relates to the transition momentum $k_\mu = p_i - p_j$, roughly of the order of $m_{d_i}$, its contributions to $d_i\to d_j \ell' \overline{\ell'}$ are indeed suppressed by $c_\theta\, m_{d_i} {\cal O}(m_{d_i})/m^2_T$ and $s_\theta {\cal O}(m_{d_i})/m_T$. Hence, we can neglect the effects of the tensor operators.

Furthermore, when all Feynman diagrams, including self-energy diagrams, are considered to cancel the ultraviolet divergences, the dominant effects of Fig.~\ref{fig:Z_penguin} arise from the vector currents. Retaining the same chirality in the initial and final quarks requires a double mass insertion in the $T$-quark propagator;  thus, the suppression factor of $1/m^2_T$ is replaced by the factor of $m^2_T$ in the induced effective coefficients. As a result, the effective Hamiltonian for $d_i \to d_j \ell \overline{\ell'}$ can be derived as:
\begin{align}
{\cal H}^{\rm NP}_{d_i \to d_j \ell' \overline{\ell'}} & = - \sqrt{2} \, G_F \, \overline{d_j} \gamma_\mu \left( C^L_{Zji} P_L + C^R_{ji} P_R \right)d_i \, \overline{\ell'} \gamma^\mu \left( c^{\ell'}_V - c^{\ell'}_A \gamma_5 \right) \ell' \,, \label{eq:Hditodj}
\end{align}
where $c^{\ell'}_V=I^3_{\ell'} - 2 s^2_W Q_{\ell'}$ and  $c^{\ell'}_A= I^3_{\ell}$ with $I^3_{\ell'}$ being the third weak isospin of $\ell'$, and the induced coefficients $C^{L,R}_Z$ and loop integral function $J_Z$ are given as:
\begin{align}
\begin{split}
C^L_{Zji} & = \frac{2 Q_T s^2_W c^2_\theta}{16 \pi^2} Y^{R}_{Tj} Y^{R*}_{Ti} J_Z\left(z_{H^+_1} \right)
\,, \\
C^R_{Zji} & = \frac{2 Q_T s^2_W c^2_\theta}{16 \pi^2} Y^{L}_{Tj} Y^{L*}_{Ti} J_Z\left(z_{H^+_2} \right)
\,, \\
J_Z(x) &= \frac{1}{1-x} + \frac{x \ln x}{(1-x)^2}\,.
\end{split}
\end{align}
When $x\to 0$ and $x\to 1$, the asymptotic values of $J_Z(x)$ are $1$ and $1/2$, respectively. Therefore, 
if we take the scheme of $s_\theta \ll 1$, the primary free parameters in the matrices ${\bf C}^{L,R}_Z$ are ${\bf Y}^{L,R}_T$.

\section{Observables in rare $B$ and $K$ decays} \label{sec:obs}

Based on the $Z$-mediated interactions shown in Eq.~(\ref{eq:Hditodj}), we discuss their contributions to the rare $B$ and $K$ decays in this section. The processes of interest inlcude $B\to K^{(*)} \nu \bar\nu$, $K^+ \to \pi^+ \nu \bar\nu$, $K_L\to \pi^0 \nu \bar\nu$, $B_s\to \mu^- \mu^+$, and  inclusive $B\to X_s \ell^- \ell^+$ decays. Since the influence on the angular observables of $B\to K^* \ell^- \ell^+$ is found to be insignificant for the parameters that enhance the BR of $B^+\to K^+ \nu \bar\nu$, we do not discuss them in this study.

\subsection{$B\to (K, K^*)  \nu \bar\nu$}

Combining Eq.~(\ref{eq:Hditodj}) with the interactions  in the SM,  the effective Hamiltonian for $b\to s \nu \bar\nu$ is written as:
 \begin{equation}
 \begin{split}
 {\cal H}_{ b\to s \nu \bar\nu}= & \frac{4 G_F}{\sqrt{2}} \frac{\alpha}{2\pi s^2_W} \left[ \left( V_{ts}^* V_{tb} X_t -\frac{\sqrt{2} G_F C^L_{Z23}}{C^\nu_{\rm SM}} \right) \bar s \gamma^\mu P_L b \bar\nu \gamma_\mu P_L \nu \right.  \\
 & \left.  - \frac{\sqrt{2} G_F C^R_{Z23}}{C^\nu_{\rm SM}} \bar s \gamma^\mu P_R b \bar\nu \gamma_\mu P_L \nu \right]~,
 \end{split}
 \end{equation}
where the neutrino flavors are suppressed, $\alpha=e^2/4\pi$, $X_t=1.481\pm 0.009$~\cite{Buras:2015qea},  and $C^{\nu}_{\rm SM}= 2 G_F \alpha/(\sqrt{2} \pi s^2_W)$.  The $q^2$-dependent differential decay rate of $B\to K \nu \bar\nu$ is obatined as:
 \begin{equation}
 \frac{d\Gamma}{d q^2} (\bar B\to K \nu \bar\nu) =  \frac{d\Gamma^{\rm SM}}{d q^2} (B\to K \nu \bar\nu)   \left|1 - \frac{\sqrt{2} G_F (C^L_{Z23}+C^R_{Z23}) }{V^*_{ts} V_{tb} X_t C^\nu_{\rm SM}} \right|^2~.
 \end{equation}
Since the new interactions are lepton flavor-conserving and involve only the vector currents, we can factorize the new physics effects as a $q^2$-independent scalar factor. Therefore, the ratio of BR in the model to the SM prediction can be simplified as: 
 \begin{equation}
 R^\nu_{K}= \frac{{\cal B}(B^+\to K^+ \nu \bar\nu)}{{\cal B}(B^+\to K^+ \nu \bar\nu)^{\rm SM}} =   \left|1 - \frac{\sqrt{2} G_F (C^L_{Z23}+C^R_{Z23}) }{V^*_{ts} V_{tb} X_t C^\nu_{\rm SM}} \right|^2~. \label{eq:RnuK}
 \end{equation}

Using the form factors of $\bar B \to K^*$ defined in Appendix~\ref{app:FFs}, the partial differential decay rate for $B\to K^* \nu \bar\nu$ is
 \begin{equation}
 \begin{split}
 \frac{d\Gamma}{d q^2} (B\to K^* \nu \bar\nu) = & \frac{G^2_F \alpha^2}{256 \pi^5 s^4_W m^3_B} q^2 \sqrt{\lambda_{K^*}}    \left[ \left| V^*_{ts} V_{tb} X_t +  \frac{\sqrt{2} G_F  }{C^\nu_{\rm SM}} (C^L_{Z23}-C^R_{Z23})\right|^2 H^2_{V,0} \right. \\
& \left. + \left( \left| V^*_{ts} V_{tb} X_t + \frac{\sqrt{2} G_F C^L_{Z23}}{C^\nu_{\rm SM}} \right|^2 + \left| \frac{\sqrt{2} G_F C^R_{Z23}}{C^\nu_{\rm SM}}\right|^2 \right) \left( H^2_{V,+} + H^2_{V-}\right) \right. \\
& \left. - 4 \operatorname{Re}\left( \left(V^*_{ts} V_{tb} X_t + \frac{\sqrt{2} G_F C^L_{Z23}}{C^\nu_{\rm SM}} \right) \frac{\sqrt{2} G_F C^R_{Z23}}{C^\nu_{\rm SM}} \right) H_{V,+} \, H_{V,-}\right]\,,
 \end{split}
 \end{equation}
where the $q^2$-dependent $\lambda_{K^*}$ and polarization factors $H_{V,0(\pm)}$ of $K^*$ are defined as~\cite{Browder:2021hbl}:
  \begin{subequations} 
  \begin{align}
  \lambda_P (q^2)& = m^4_B + m^4_{P} + q^4 -2 m^2_B m^2_P -2 m^2_P q^2 -2 m^2_B q^2 
  \,, \\
 H_{V,0}(q^2) & = \frac{8}{\sqrt{q^2}} m_B m_{K^*} A_{12} (q^2)
 \,, \\
 H_{V,\pm}(q^2) & =  (m_B + m_{K^*}) A_1(q^2) \mp \frac{\sqrt{\lambda_{K^*}}}{m_B + m_{K^*} } V(q^2)\,,
  \end{align}
 \end{subequations} 
with $A_{12}(q^2)$, $A_1(q^2)$, and $V(q^2)$ being the $\bar B\to K^*$ transition form factors.  In our numerical estimates, we use the $\bar B \to K$ form factors obtained from the lattice QCD calculations~\cite{Bailey:2015dka,Parrott:2022rgu}, and the $\bar B\to K^*$ form factors obtained from the combination of light-cone sum rule and lattice QCD calculations~\cite{Bharucha:2015bzk}.  To illustrate the influence of new physics, we define analogous to Eq.~\eqref{eq:RnuK} the ratio:
  \begin{equation}
 R^\nu_{K^*}= \frac{{\cal B}(B_d\to K^{*0} \nu \bar\nu)}{{\cal B}(B_d\to K^{*0} \nu \bar\nu)^{\rm SM}} ~. \label{eq:RnuKv}
 \end{equation}

\subsection{$K^+ \to \pi^+ \nu \bar\nu $ and $K_L \to \pi \nu \bar\nu$}

 The effective Hamiltonian for $d\to s \nu \bar\nu$ in the model is written as:
 \begin{align}
 \begin{split}
 {\cal H}_{d \to s \nu \bar\nu} = & C^\nu_{\rm SM}  \sum_{\ell =e, \mu \tau} \left[ \left( V^*_{cs} V_{cd} X^\ell_{\rm NNL} + V^*_{ts} V_{td} X_t - \frac{\sqrt{2} G_F C^L_{Z21} }{C^\nu_{\rm SM}} \right) \bar s \gamma^\mu P_L d \right. \\
 & \left. \qquad\qquad\qquad
 -  \frac{\sqrt{2} G_F C^R_{Z21} }{C^\nu_{\rm SM}} \bar s \gamma^\mu P_R d  \right] \bar\nu_\ell  \gamma_\mu P_L \nu_\ell ~,
 \end{split}
 \end{align}
where $X^\ell_{\rm NNL}$ denotes the contributions from the charm quark with the QCD corrections calculated up to next-to-next-to-leading (NNL) order~\cite{Buras:2005gr,Buras:2006gb}, and the two-loop electroweak corrections also included~\cite{Brod:2008ss}.  Using the results formulated in Refs.~\cite{Mescia:2007kn,Buras:2015qea},  the BRs of the $K^+\to \pi^+ \nu \bar\nu$ and $K_L\to \pi^0 \nu \bar\nu$ decays can be obtained respectively as:
\begin{align}
\begin{split}
{\cal B}(K^+ \to \pi^+ \nu \bar\nu) 
&=  \kappa_+ (1+\Delta_{\rm EM})  \left[ \left( \frac{{\rm Im} X_{\rm eff} }{\lambda^5} \right)^2 
+ \left( \frac{{\rm Re}(V^*_{cs} V_{cd} )}{\lambda} P_c (X)+ \frac{{\rm Re}(X_{\rm eff})}{\lambda^5}\right)^2 \right]
~, \\
{\cal B}(K_L \to \pi^0 \nu \bar\nu) 
& ={\cal B}(K_L \to \pi^0\nu \bar\nu)^{\rm SM}    \left|1 - \frac{\sqrt{2} G_F \operatorname{Im}(C^L_{Z21} + C^R_{Z21})}{ \operatorname{Im}(V^*_{ts} V_{td})X_t C^\nu_{\rm SM}} \right|^2 
~, 
\end{split}
\label{eq:KpL}
\end{align}
where $\Delta_{\rm EM} = -0.003$ is the electromagnetic radiative corrections, and $\kappa_+ =( 5.173 \pm 0.025)\times 10^{-11} (\lambda/0.225)^8$.  Here $P_c(X) = P^{\rm SD}_c (X) + \delta P_{c,u}=0.405\pm 0.024$ denotes the charm-quark loop contributions, in which the short-distance part is given by~\cite{Buras:2015qea}:
 \begin{equation}
 P^{\rm SD} _c(X) = \frac{1}{\lambda^4} \left( \frac{2}{3} X^e_{\rm NNL} + \frac{1}{3} X^\tau_{\rm NNL}\right)=0.365\pm 0.012\,,
 \end{equation}
 and the long-distance contribution is estimated as $\delta P_{c,u}=0.04\pm 0.02$~\cite{Isidori:2005xm}.  The factor $X_{\rm eff}$  that combines the SM and new physics effects is:
 \begin{equation}
 X_{\rm eff} = V^*_{ts} V_{td} X_t - \frac{\sqrt{2} G_F}{C^\nu_{\rm SM}} (C^L_{Z21} + C^R_{Z21}) \,. \label{eq:xeff}
 \end{equation}
Similar to $R^{\nu}_{K, K^*}$, we will explore the new physics effects contributing to rare $K$ decays by using the ratio of the BR in the model to the SM prediction, defined as:
 \begin{equation}
 R^{\nu}_{\pi} = \frac{{\cal B}(K\to \pi \nu \bar\nu)}{{\cal B}(K\to \pi \nu \bar\nu)^{\rm SM}}\,, \label{eq:Rnupi}
 \end{equation}
where $\pi=\pi^+ (\pi^0)$ when $K=K^+(K_L)$.

It is worth mentioning that the new physics effects on $B^+\to K^+ \nu\bar\nu$ and $K_L\to \pi^0 \nu \bar\nu$ can be factorized as a multiplicative factor as shown in Eqs.~(\ref{eq:RnuK}) and (\ref{eq:KpL}). Since we adopt $V^d_R=V_{\rm CKM}$ in this study, the multiplicative factor in both equations indeed is approximately the same when the Yukawa couplings follow the hierarchy of $y^{R(L)}_{T1} \ll y^{R(L)}_{T2} \ll y^{R(L)}_{T3}$ due to the constraints from the $|\Delta F| = 2$ processes. If the small CP phase of  $V_{ts}$ is neglected, the new physics effect in Eqs.~(\ref{eq:RnuK}) and (\ref{eq:KpL}) can be expressed as:
 \begin{align}
 \begin{split}
 \frac{ (C^L_{Z23} +C^R_{Z23})}{V^*_{ts} V_{tb} } 
 & \approx  
 \frac{ \operatorname{Im}(C^L_{Z21} +C^R_{Z21})}{Im(V^*_{ts} V_{td}) }  
 \approx \frac{2Q_T s^2_W c^2_\theta}{16\pi^2} \left[ \left(y^R_{T3} + \frac{y^R_{T2}}{V_{ts}}\right)y^R_{T3} + R \to L\right]\,. \label{eq:RnuK=Rnupi0}
 \end{split}
 \end{align}
 That is, we obtain $R^\nu_K \simeq R^\nu_{\pi^0}$ in the model. We will explicitly demonstrate the relationship in the numerical analysis.

\subsection{$B\to X_s \ell^- \ell^+$ and $B_s \to \mu^- \mu^+$}

 The effective Hamiltonian for the inclusive and exclusive $b\to s \ell^- \ell^+$  decays, including the loop matrix element effects that arise from the tree- and penguin-induced four-quark operators, is given as:
 \begin{equation}
 {\cal H}_{b\to s \ell^- \ell^+} = - \frac{4 G_F}{\sqrt{2}} \frac{\alpha}{4 \pi} V^*_{ts} V_{tb} \left[ \sum_{k=9,10} \left(C^{\rm eff}_k {\cal O}_k + C^{\prime \rm eff}_k {\cal O}'_k \right) - \frac{2m_b}{q^2} C^{\rm eff}_7 {\cal O}_7 \right]\,. \label{eq:Hbsll}
 \end{equation}
The effective operators in the model are 
  \begin{subequations}
 \begin{align}
 {\cal O}^{(\prime)}_9 & = \left(\bar s  \gamma_\mu P_{L(R)}   b \right)  \left(\bar \ell \gamma^\mu \ell \right)
 ~, \\
   {\cal O}^{(\prime)}_{10} & = \left(\bar s  \gamma_\mu P_{L(R)}  b \right)  \left(\bar \ell \gamma^\mu \gamma_5 \ell \right) 
   ~, \\
 {\cal O}_{7} & = \left(\bar s  i \sigma_{\mu \nu} q^\nu  P_R  b \right) \left(\bar \ell \gamma^\mu  \ell \right)~, 
 \end{align}
  \end{subequations}
 and  the effective coefficients, combining the SM contributions and the effects from the new $Z$-penguin diagrams, are given as:
 \begin{subequations}
 \begin{align}
 C^{\rm eff}_9 & = C^{\rm eff, SM}_9 + \frac{\sqrt{2} G_F }{C^\ell_{\rm SM}} c^\ell_V C^L_{Z23} 
 ~, \\
 C^{\prime \rm eff}_9 & = \frac{\sqrt{2} G_F }{C^\ell_{\rm SM}} c^\ell_V C^R_{Z23} 
 ~, \\
 C^{\rm eff}_{10} & = C^{\rm eff, SM}_{10} - \frac{\sqrt{2} G_F }{C^\ell_{\rm SM}} c^\ell_A C^L_{Z23} 
 ~, \\
 C^{\prime \rm eff}_{10} & =- \frac{\sqrt{2} G_F  }{C^\ell_{\rm SM}} c^\ell_A C^R_{Z23} ~,
 \end{align}
  \end{subequations}
  with $C^{\ell}_{SM}= G_F \alpha V^*_{ts} V_{tb} /(\sqrt{2} \pi)$. 
The SM results, obtained by ignoring the $q^2$-dependence at the energy scale of $\mu_b=4.2$ GeV, are $C^{\rm eff,SM}_9\approx 4.114$, $C^{\rm eff, SM}_{10}\approx -4.193$, and $C^{\rm eff}_7=C^{\rm eff,SM}_7 \approx -0.2957$~\cite{Blake:2016olu}. Their detailed NNLO results can be found in Refs.~\cite{Bobeth:1999mk,Bobeth:2003at,Huber:2005ig}.

Since the new $Z$-mediated contributions to the angular observables of $B\to K^* \ell^- \ell^+$ are insignificant for the effects that enhance the $d_i \to s \nu \bar\nu$ processes, our attention focuses on the contributions to the $B\to X_s \ell^- \ell^+$ and  $B_s \to \mu^- \mu^+$ processes, where both processes play a crucial role in constraining the parameters related to the $b\to s$ transition. Applying the interactions in Eq.~(\ref{eq:Hbsll}), the differential decay rate for $b\to X_s \ell^- \ell^+$ as a function of $s=q^2/m^2_b$ can be obtained as:
\begin{align}
\begin{split}
\frac{\Gamma(b\to X_s \ell^- \ell^+)}{ds} 
=&
\left( \frac{\alpha}{4 \pi} \right)^2  \frac{G^2_F m^5_b }{48 \pi^3}  \left| V^*_{ts} V_{tb}\right|^2 (1-s)^2 \Bigg[ \left( 4 + \frac{8}{s}\right) |C^{\rm eff}_7|^2 +(1+2 s) \left( |C^{\rm eff}_{9}|^2 \right. \\
&  \left. 
+|C^{\rm \prime eff}_{9}|^2  + |C^{\rm eff}_{10}|^2 +|C^{\rm \prime eff}_{10}|^2\right) +12 C^{\rm eff}_{7} Re(C^{\rm eff}_9 + C^{\prime \rm eff}_9) \Bigg]~.
\end{split}
\end{align}
Notably, ${\cal O}^{(\prime)}_9$ and ${\cal O}_7$ do not contribute to the chirality suppression process $B_s \to \mu^- \mu^+$. 
The BR for $B_s\to \mu^- \mu^+$ arising from ${\cal O}^{(\prime)}_{10}$ is given as~\cite{DeBruyn:2012wk,Fleischer:2012fy,Buras:2013uqa}:
\begin{align}
{\cal B} (B_s \to \mu^- \mu^+)  =\tau_{\mu^- \mu^-}  \frac{G^2_F}{\pi} \left| \frac{\alpha V^*_{ts} V_{tb}}{4 \pi}\right|^2 F^2_{B_s} m_{B_s} m^2_\mu \sqrt{1- \frac{4 m^2_\mu}{m_{B_s}}} \left| C^{\rm \prime}_{10} -C^{\rm}_{10}  \right|^2\,,
\end{align}
where $\tau_{\mu^- \mu^+}$ is the effective lifetime of $B_s$ in the time-dependent $B_s \to \mu^- \mu^+$ decay and is related to the width difference between heavy and light $B_s$ mesons. The simplified relation can be written as $\tau_{\mu^- \mu^+}=\tau_{B_s}/(1-y_s)$ with $2 y_s=\tau_{B_s} \Delta\Gamma_s =0.128 \pm 0.007$~\cite{HFLAV:2022esi} and $\tau_{B_s}$ being the $B_s$ lifetime. The Wilson coefficients $C^{(\prime)}_{10}$ are $C'_{10}=C^{\rm \prime eff}_{10}$ and 
 \begin{equation}
C_{10}= \eta_{\rm eff} \frac{0.315 x^{0.78}_t}{s^2_W}   - \frac{\sqrt{2} G_F }{C^\ell_{\rm SM}} c^\ell_A C^L_{Z23}~,
\end{equation}
where $\eta_{\rm eff}=0.9882\pm 0.00024$ represents the QCD and electroweak corrections~\cite{Bobeth:2013uxa,Buras:2013dea}.

\section {Numerical analysis and discussions} \label{sec:num}

There are only seven observables measured from the neutrino oscillation experiments. Therefore, the free parameters in the one-loop induced neutrino mass matrix given in Eq.~(\ref{eq:nu_mass}), such as $m_{S_{1,2}, N}$, $\phi$, and $y_{1i,2i}$, cannot be completely fixed. More related processes, such as $\ell_i \to \ell_j \gamma$, $\mu\to 3e$, $\mu-e$ conversion, and muon $g-2$, should be included and analyzed together. Since the $m_N$, $\phi$, and $y_{1i,2i}$ parameters are irrelevant to the semileptonic $B$ and $K$ decays,  we don't repeat the numerical analysis of the neutrino physics and lepton flavor-violating processes in this work. The relevant study with detailed analysis can be found in Ref.~\cite{Chen:2022gmk}. Hence, we focus on the contributions from the parameters ${\bf y}^{R,L}_T$, $m_{H^+_{1,2}}$ and their mixing angle $\theta$, and $m_T$.

\subsection{Numerical inputs and parameter constraints}

As discussed earlier, the weak phases of ${\bf y}^{R,L}_{T}$ can be rotated away; therefore, six free parameters are involved in the quark Yukawa couplings. To satisfy the perturbativity requirement, we assume their upper limits to be $|{\bf y}^{R,L}_{Ti}|< \sqrt{4\pi}$.  According to the fact that $\Delta m_K \ll \Delta m_{B_d} < \Delta m_{B_s}$ and $\Delta m^{\rm exp}_{B_d}/\Delta m^{\rm exp}_{B_s} \sim \Delta m^{\rm SM}_{B_d}/\Delta m^{\rm SM}_{B_s} \sim \lambda^2$ with $\lambda\approx 0.225$ being one of the Wolfenstein parameters, we further restrict the upper bounds on the parameters to be $|y^{R,L}_{T3}|\lesssim 3.5$, $|y^{R,L}_{T2}|\lesssim 1$, and $|y^{R,L}_{T1}|\lesssim 0.5$ in the numerical calculations.

For the mass limit of the heavy quark $T$, we adopt the constraints based on the stop searches in $R$-parity conserving supersymmetry (SUSY).  The data, obtained with an integrated luminosity of 139~fb$^{-1}$ at $\sqrt{s}=13$~TeV by ATLAS~\cite{ATLAS:2021hza}, show that mass below $1$~TeV has been excluded when the neutralino mass is around 100~GeV. Since the lightest neutral inert scalar cannot be the DM in the model, i.e., $m_{S_{1,2}} > m_N$, the inert charged scalar should also be heavier than $N$. Thus, to reduce the number of free parameters in our numerical analysis, we fix the masses as follows: $m_T=1.2$~TeV, $m^2_{H^+_1}/m^2_T=0.3$, and $m^2_{H^+_2}/m^2_T=0.4$. For the scenario with a small mixing angle $\theta$, we fix $s_\theta=0.1$ in the numerical estimates.

With the above-specified values of $m_T$ and $m_{H^+_{1,2}}$, more than six experimental observables are required to determine the allowed ranges of ${\bf y}^{R,L}_{T}$. Since we want to consider the processes of $B\to K^{(*)} \nu \bar\nu$ and $K_L\to \pi^0 \nu \bar\nu$ as predictions in the model, they should be excluded from the inputs. Based on the discussions in Secs.~\ref{sec:processes} and \ref{sec:obs}, the experimental inputs can be $\Delta m_{K, B_q}$, $B_s \to \mu^- \mu^+$, $B\to X_s  \ell^- \ell^+$, $K^+\to \pi^+ \nu \bar \nu$ as well as the CP asymmetry in the $B\to J/\Psi K$ decay, denoted as $S_{J/\Psi K^0}$. Thus, we can use these seven observables to put bounds on the six free parameters. Their SM predictions and current experimental values are shown in Table~\ref{tab:inputs}. The SM results for $B\to K^{(*)} \nu \bar \nu$ used in our numerical estimates are~\cite{Buras:2022qip} 
 \begin{align}
 {\cal B}(B^+\to K^+ \nu \bar\nu) & =(4.92 \pm 0.30)\times 10^{-6}\,, \nonumber \\
 {\cal B}(B_d \to K^{*0} \nu \bar\nu) & =(10.13 \pm 0.92)\times 10^{-6}\,.
 \end{align}
As a comparison, the lattice QCD results obtained in Ref.~\cite{Becirevic:2023aov} are ${\cal B}(B^+\to K^+ \nu \bar\nu) =(5.06\pm0.14\pm0.28)\times 10^{-6}$ and ${\cal B}(B_d \to K^{*0} \nu \bar\nu) = (9.05\pm1.25\pm 0.55)\times 10^{-6}$.

\begin{table}[thp]
\caption{The experimental measurements and the SM predictions.}
\begin{center}
\begin{tabular}{c|cccc}
\hline \hline
 Obs. & ~$\Delta m_{K}\cdot 10^{15}$ [GeV]~  & ~$\Delta m_{B_d} \cdot 10^{13}$ [GeV]~& ~ $\Delta m_{B_s} \cdot 10^{12}$ [GeV]~  & ~ $S_{J/\Psi K^0}$\\ \hline
 Exp.~\cite{ParticleDataGroup:2022pth} & $3.482\pm 0.006$ & $3.332\pm 0.013$  & $11.688 \pm 0.004 $  & $0.699\pm 0.017$ 
 \\ \hline
 SM & $5.8 \pm 2.4$~\cite{Wang:2022lfq}   & $3.618^{+1.052}_{-0.987}$~\cite{Lenz:2010gu}   & $11.053^{+3.618}_{-2.237}$~\cite{Lenz:2010gu}   & $0.831\pm 0.116$~\cite{Lenz:2010gu} 
 \\  \hline \hline 
 Obs.  & ~${\cal B}(B_s \to \mu^- \mu^+) \cdot 10^{9}$~ & ~${\cal B}(b\to X_s \ell^- \ell^+)\cdot 10^6$~ &~ ${\cal B}(K^+ \to \pi^+ \nu \bar\nu)\cdot 10^{11}$ ~& ~${\cal B}(K_L \to \pi^0 \nu \bar\nu) \cdot 10^{11}$
 \\ \hline
 Exp. &  $3.01\pm 0.35$~\cite{ParticleDataGroup:2022pth}  & $5.8 \pm 1.3$~\cite{ParticleDataGroup:2022pth} &  $11.4^{+4.0}_{-3.3}$~\cite{ParticleDataGroup:2022pth}  &  $< 2 \times 10^{-9}$~\cite{KOTO2023} 
 \\ \hline 
 SM  & $3.78^{+0.15}_{-0.10}$~\cite{Buras:2022qip} & $4.15 \pm 0.7$~\cite{Bobeth:2007dw}  &  $8.60 \pm 0.42$~\cite{Buras:2022qip} & $2.94\pm 0.15$~\cite{Buras:2022qip}
 \\ \hline\hline
\end{tabular}
\end{center}
\label{tab:inputs}
\end{table}

To determine the ranges of the free parameters that are consistent with the chosen experimental data, we employ the minimum chi-square approach, where the weighted $\chi^2$ is defined as follows:
 \begin{equation}
 \chi^2 = \sum_i \frac{ (O^{\rm th}_i -O^{\rm exp}_i )^2} {\sigma^2_i}\,. \label{eq:chi2}
 \end{equation}
Here $O^{\rm th}_i $ and $O^{\rm exp}_i$ represent the central values of the $i$-th observable predicted by the model and measured in experiments, respectively. The weight factor $\sigma^2_i =( \sigma^{\rm SM}_i)^2 + (\sigma^{\rm exp}_i)^2$~\cite{Descotes-Genon:2012isb} combines the uncertainties from both SM predictions and experimental data.

With the assumed values of masses and ranges of Yukawa couplings, the minimum value of the weighted $\chi^2$ for the seven observable inputs is $\chi^2_{\rm min}=0.23$, while the $\chi^2$ value in the SM is $\chi^{2}_{\rm SM}=4.93$.  The best-fit parameter values are ${\bf y}^{L,\rm min}_{T}\simeq (0.032,\, 0.150,\, 0.893)$ and ${\bf y}^{R,\rm min}_T=(0.024, \, 0.111, \,0.813)$. To clearly understand the parameter correlations, we show contours of the $\chi^2$ function in the planes of $y^R_{T3}$-$y^{R}_{T2}$,  $y^{R}_{T3}$-$y^{R}_{T1}$, $y^{L}_{T3}$-$y^{L}_{T2}$, and $y^L_{T3}$-$y^{L}_{T1}$ in Fig.~\ref{fig:chi2}(a)-(d), where the shaded areas represent probabilities of the $\chi^2$ distribution within $68.27\%$, $95.45\%$, and $99.73\%$ confidence level (CL), respectively. Note that when two parameters are selected as variables for the two-dimensional contours, the other parameters are held fixed at their best-fit values.

\begin{figure}[phtb]
\begin{center}
\includegraphics[scale=0.38]{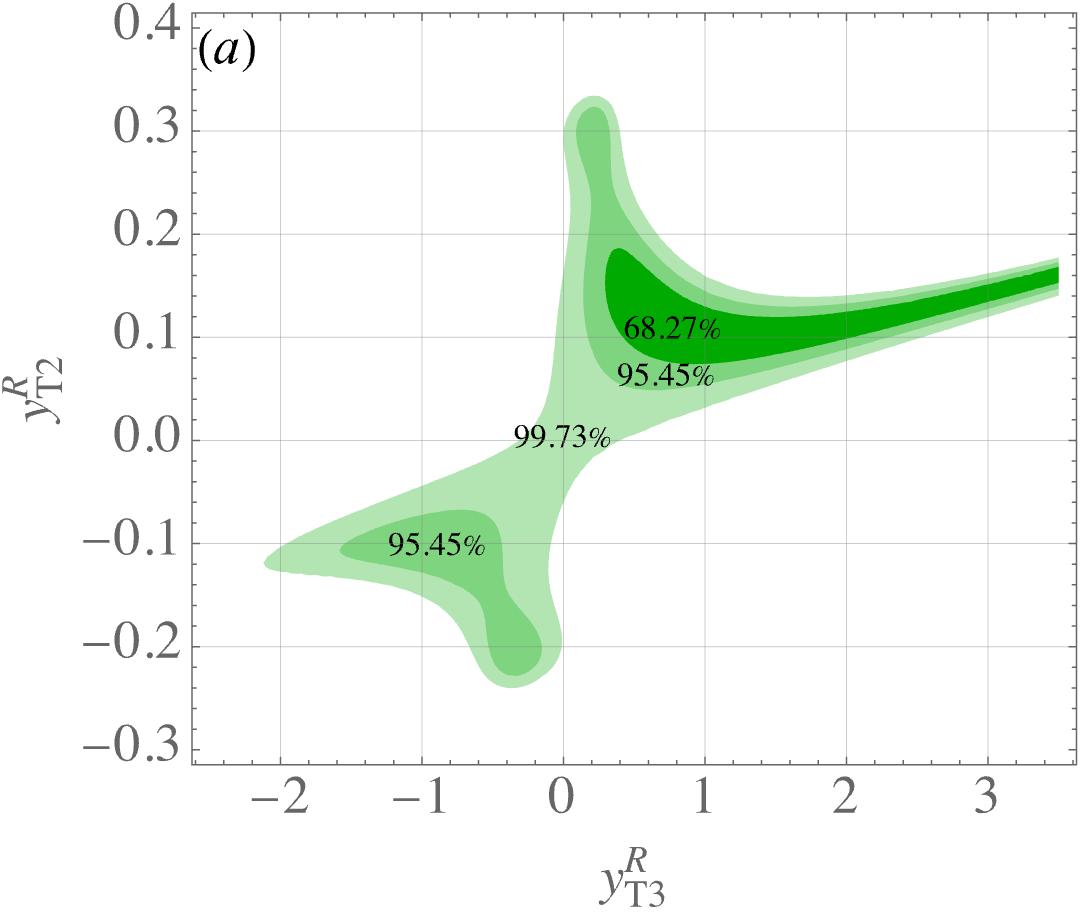}
\includegraphics[scale=0.405]{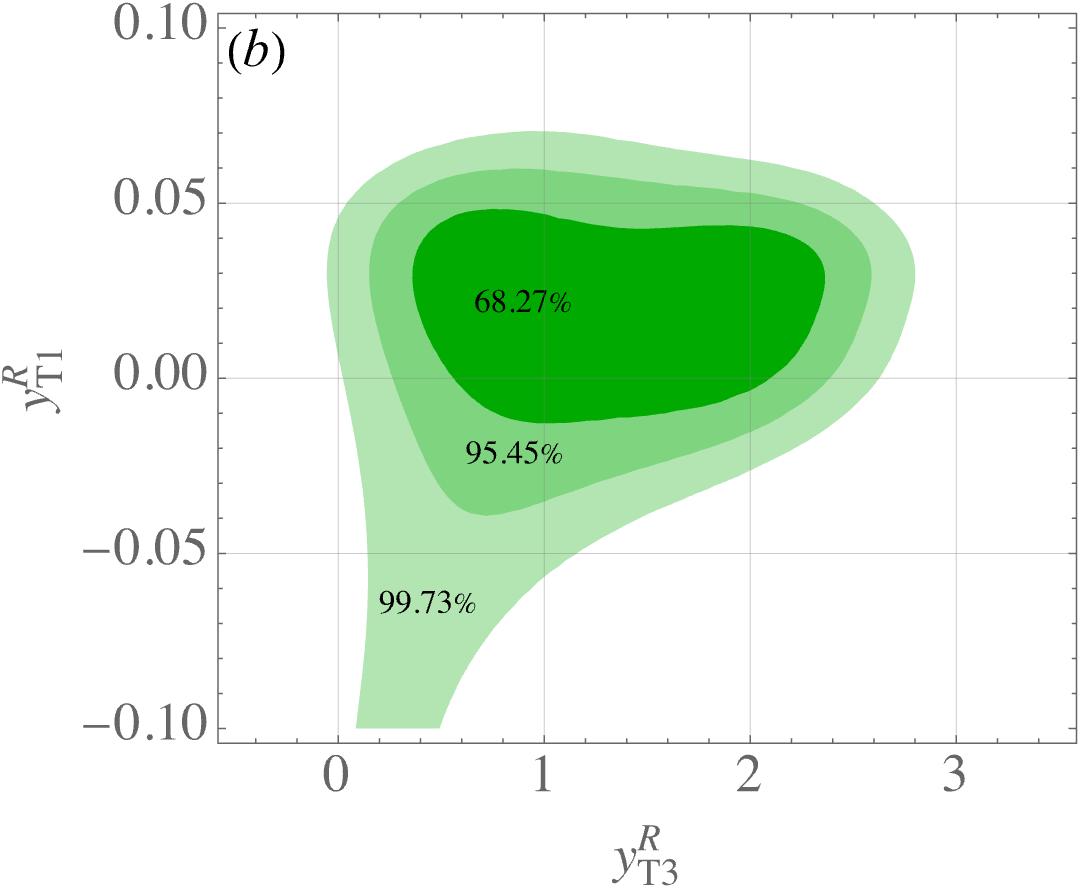}\\
\vspace{1mm}
\includegraphics[scale=0.38]{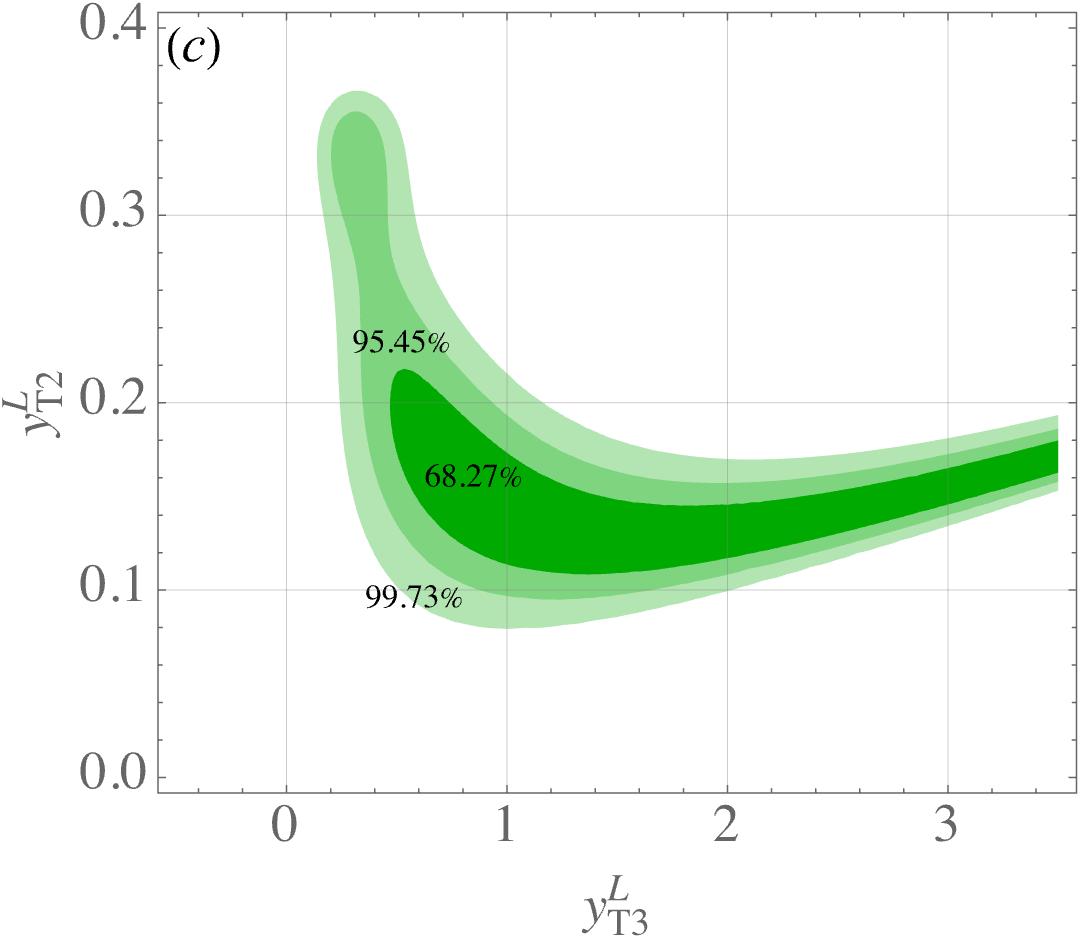} 
\includegraphics[scale=0.405]{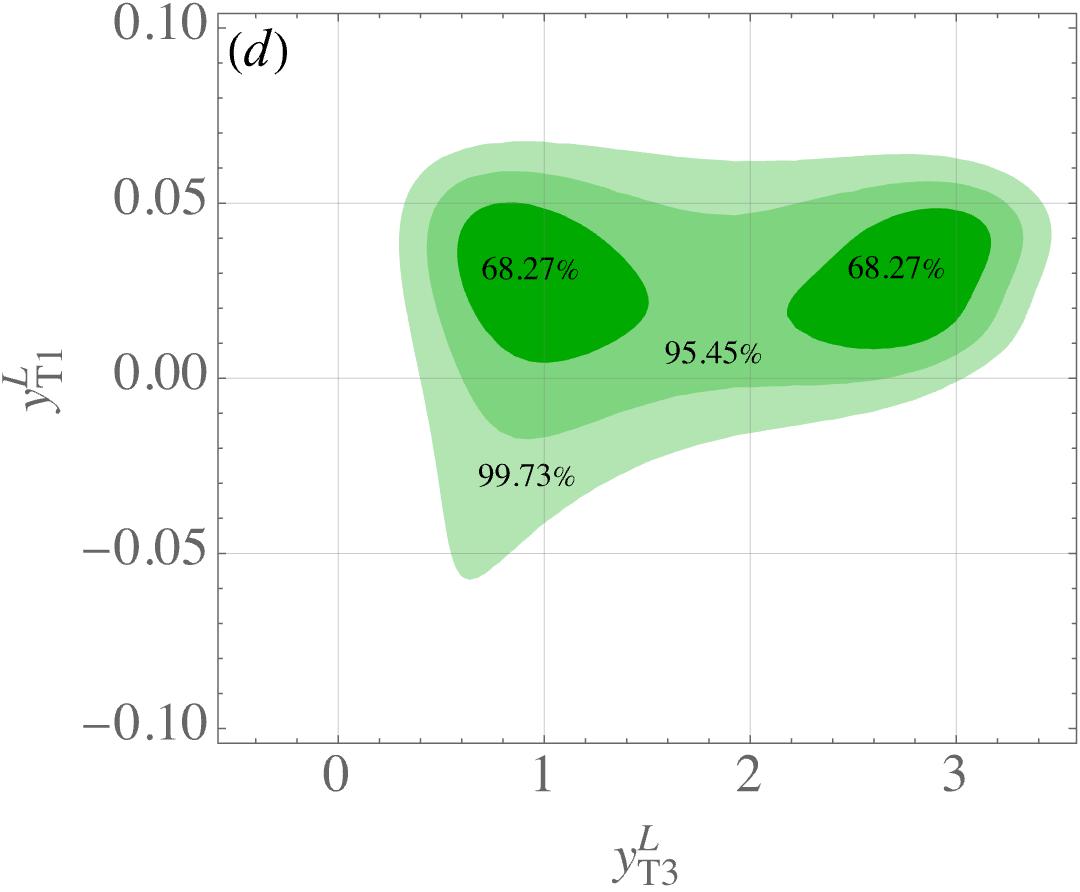}
 \caption{ Probabilities of $\chi^2$ distribution within $68.27\%$, $95.45\%$, and $99.73\%$ shown in the planes of (a) $y^R_{T3}$-$y^{R}_{T2}$, (b) $y^R_{T3}$-$y^{R}_{T1}$, (c) $y^L_{T3}$-$y^{L}_{T2}$, and (d) $y^L_{T3}$-$y^{L}_{T1}$. }
\label{fig:chi2}
\end{center}
\end{figure}

\subsection{Numerical analysis of observables in rare $B$ and $K$ processes}

 Based on the results in Figs.~\ref{fig:chi2}(b) and (d), it is seen that the ranges of $y^{R,L}_{T1}$ are localized in narrow regions at around $y^{R, \rm min}_{T1}$ and $y^{L,\rm min}_{T1}$. 
To efficiently illustrate numerical results for the studied processes in two-dimensional contour plots, we always fix $y^{R}_{T1}=y^{R, \rm min}_{T1}$ and $y^L_{T1}=y^{L,\rm min}_{T1}$. Since the areas within $2\sigma$ CL in the $y^R_{T3}$-$y^{R}_{T2}$ and $y^L_{T3}$-$y^{L}_{T2}$ planes with $y^{R,T}_{T3}>0$ have similar patterns and regions of parameters, we will demonstrate the observables in the rare $B$ and $K$ meson decays in the $y^{R}_{T3}$-$y^R_{T2}$ and $y^L_{T3}$-$y^R_{T3}$ planes. When we vary the parameters in the planes of $y^{R}_{T3}$-$y^R_{T2}$ and $y^L_{T3}$-$y^R_{T3}$ within the above-mentioned contour regions, the other parameters including $y^{R,L}_{T1}$ are fixed at ${\bf y}^{R, \rm min}_T$ and ${\bf y}^{L, \rm min}_T$.

According to $R^\nu_K$ and $R^\nu_{K^*}$ defined in Eqs.~(\ref{eq:RnuK}) and (\ref{eq:RnuKv}), the contours for $R^\nu_{K}$ (dashed) and $R^\nu_{K^*}$ (dot-dashed) in the $y^R_{T3}$-$y^R_{T2}$ plane are shown in Fig.~\ref{fig:obsyR3yR2}(a), with the values $R^\nu_{K}=(1,5,\, 1.8, \, 2.0)$ and $R^\nu_{K^*}=(0.9,\, 1.0,\, 1.1,\, 1.2)$. The results indicate that $R^\nu_K$ can reach up to a factor of 2, while the influence of new physics effects in the model on $B\to K^* \nu \bar\nu$ is mild. Additionally, we show the BR of $B_s \to \mu^- \mu^+$ within the $1\sigma$ error of the experimental value in dotted curves. As stated in the introduction, a tension exists between the experimental data and SM prediction for ${\cal B}(B_s \to \mu^- \mu^+)$. Interestingly, the $Z$-penguin diagrams mediated by the inert charged scalars can resolve this tension and significantly enhance $R^\nu_K$. For comparison, we show the $\chi^2$ contours within the $3\sigma$ CL in the plot.

\begin{figure}[phtb]
\begin{center}
\includegraphics[scale=0.4]{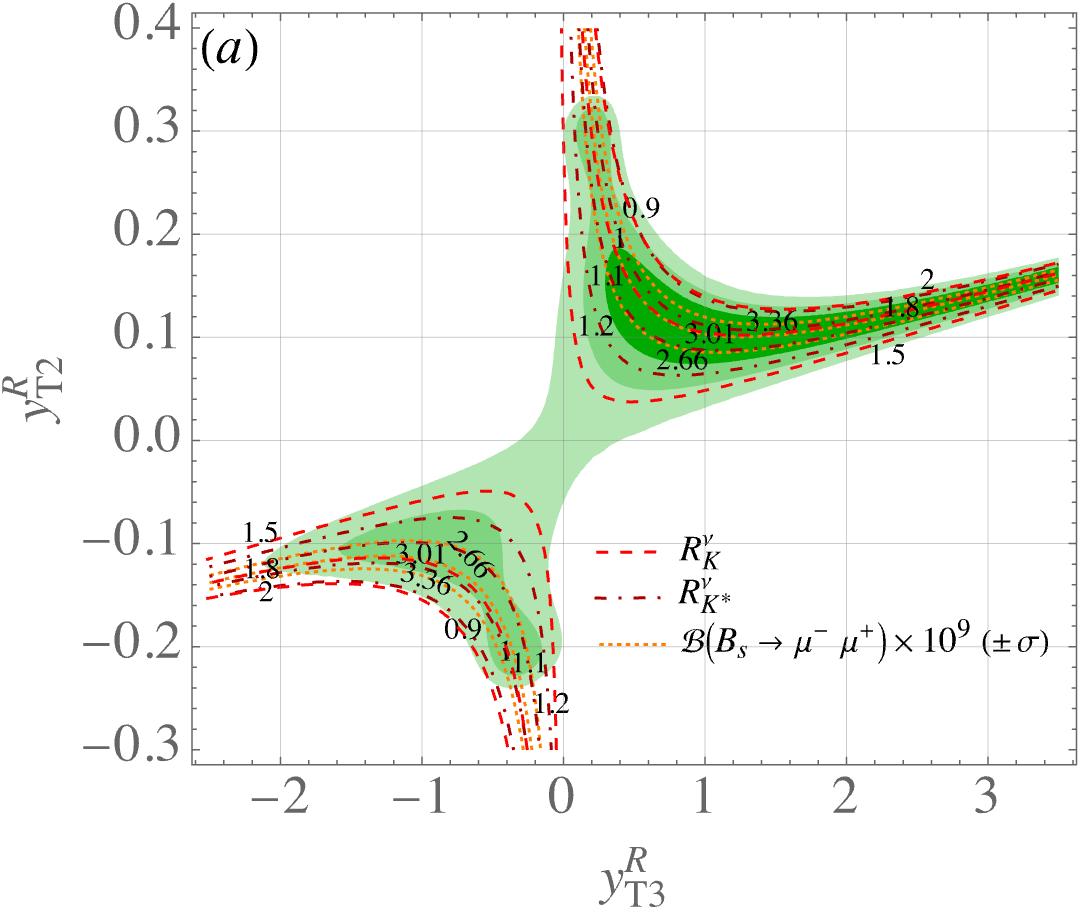}\hspace{1mm}
\includegraphics[scale=0.4]{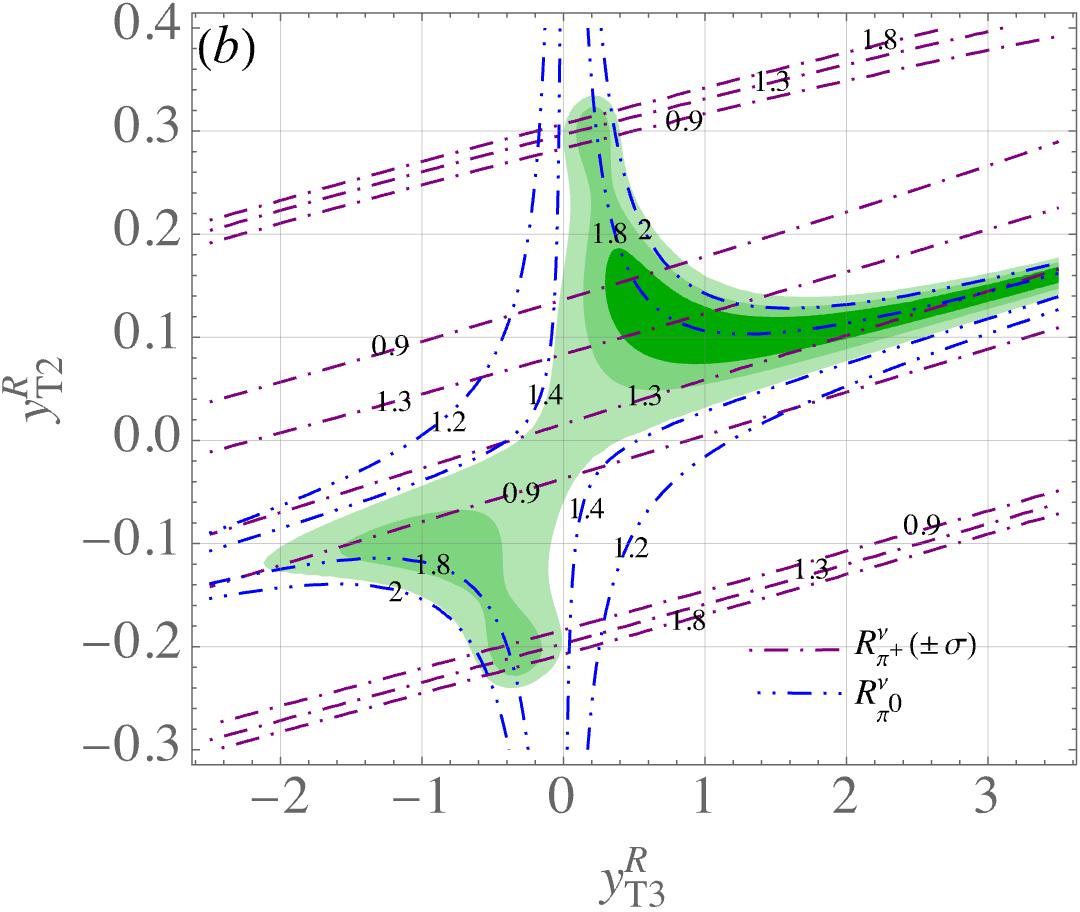}\\
\vspace{2mm} \hspace{3mm}
\includegraphics[scale=0.38]{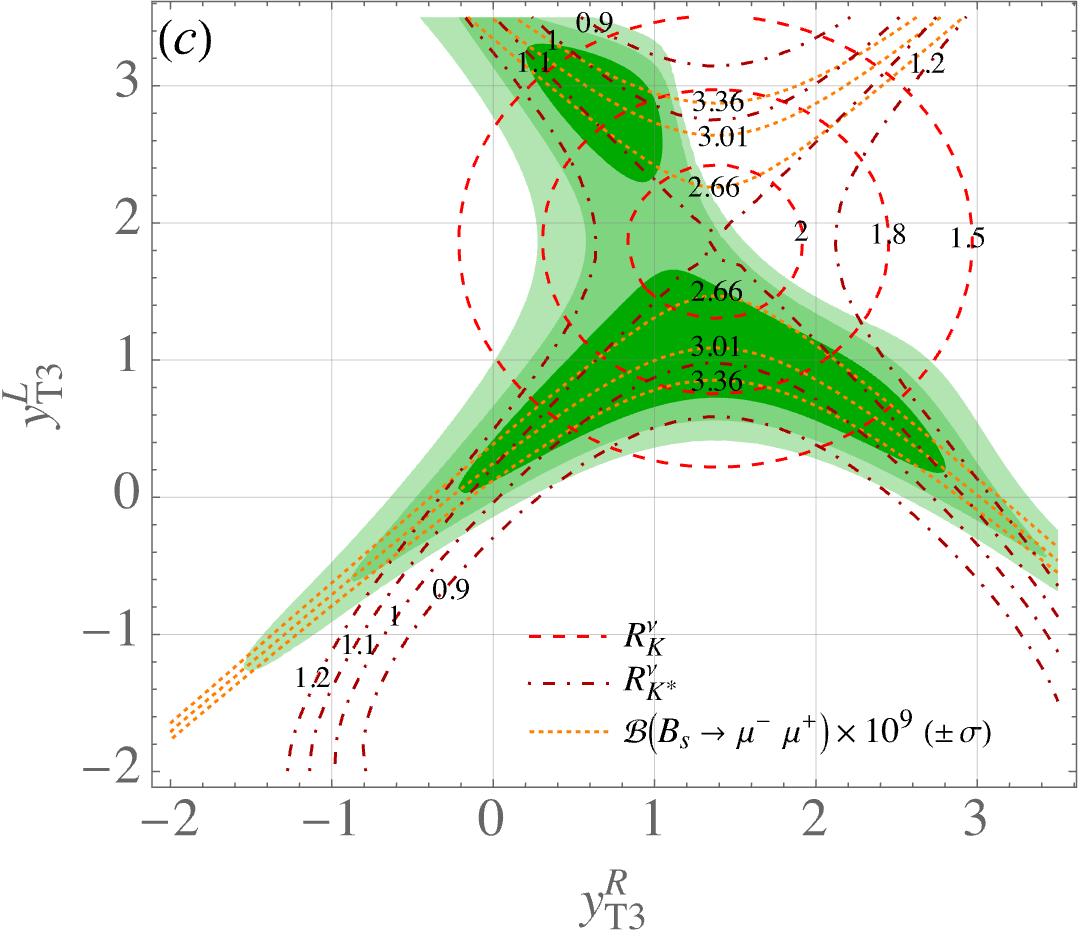} \hspace{4mm}
\includegraphics[scale=0.38]{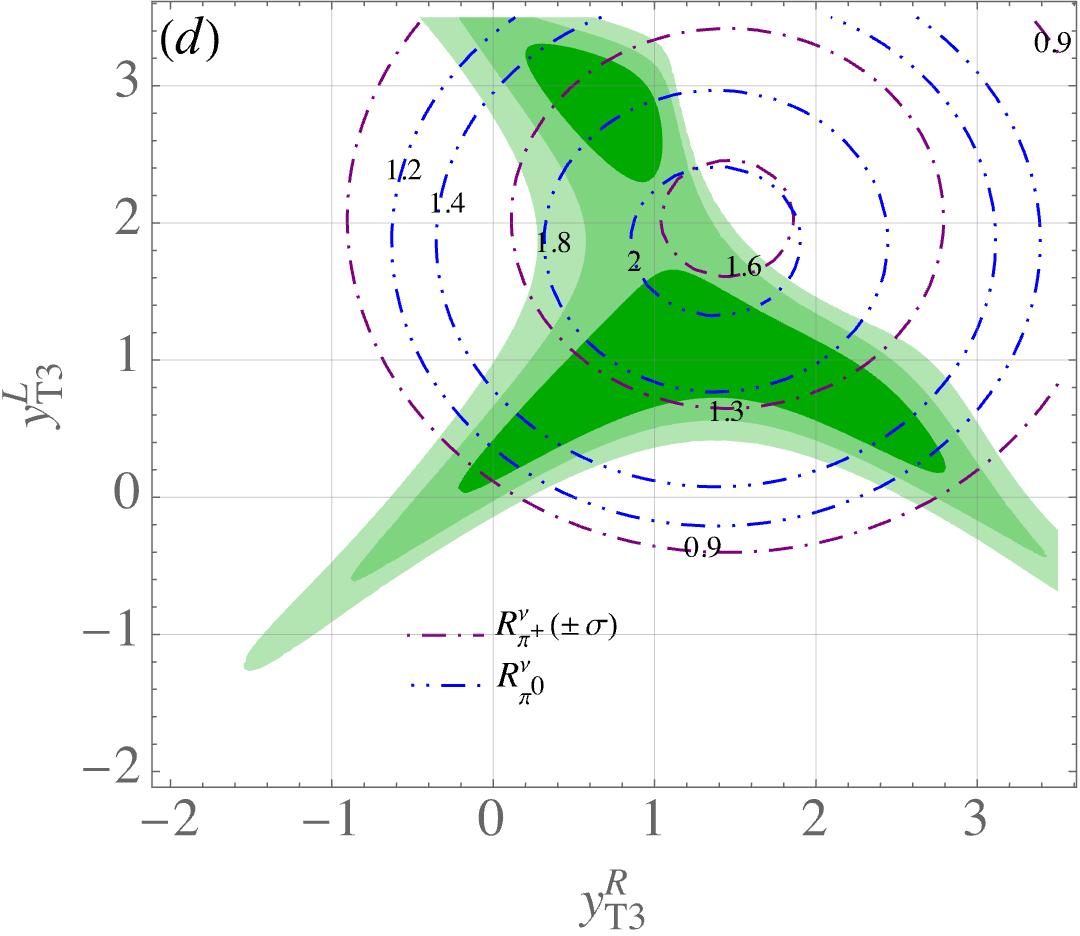}
 \caption{  Contours in the $y^R_{T3}$-$y^R_{T2}$ plane for the observables of (a) $R^\nu$, $R^\nu_{K^*}$, and ${\cal B}(B_s\to \mu^- \mu^+)$ and of (b) $R^\nu_{\pi^+}$ and $R^\nu_{\pi^0}$, where values of the parameters other than $y^R_{T3,T2}$ are taken to be their best-fit values at $\chi^2_{\rm min}$. In the plots, $\pm \sigma$ denotes that the experimental data within one standard deviation are applied. Plots (c) and (d) resemble plots (a) and (b), but in the $y^L_{T3}$-$y^R_{T3}$ plane. }
\label{fig:obsyR3yR2}
\end{center}
\end{figure}

From the definition in Eq.~(\ref{eq:Rnupi}), we show the contours for $R^\nu_{\pi^+}$ (dot-dashed) and $R^\nu_{\pi^0}$ (double-dot-dashed) in the plane of $y^R_{T3}$ and $y^{R}_{T2}$ in Fig.~\ref{fig:obsyR3yR2}(b), with the values $R^{\nu}_{\pi^+}=(0.9,\,1.3,\,1.8)$ and $R^\nu_{\pi^0}=(1.2,\, 1.4, \,1.8, \, 2.0)$, where the values of $R^{\nu}_{\pi^+}$ correspond the current experimental data with the $1\sigma$ error. The results indicate that $R^\nu_{\pi^0}=2$ can be achieved within the $3\sigma$ CL in the model, and the BR for $K^+\to \pi^+ \nu \bar\nu$ can be enahnced up to the $+1\sigma$ upper value of ${\cal B}(K^+\to \pi^+ \nu \bar\nu)\simeq 15.4\times 10^{-11}$ at $y^{R}_{T2}\sim 0.3$ and $-0.2$.

In Figs.~\ref{fig:obsyR3yR2}(c) and (d), we project the allowed parameter space onto the $y^R_{T3}$-$y^L_{T3}$ plane the same physical quantities as those shown in Figs.~\ref{fig:obsyR3yR2}(a) and (b), where $y^{L,R}_{T3}$ are the only Yukawa couplings that can be of ${\cal O}(1)$ in our setting.  Note that the contour of the $+1\sigma$ upper value $R^\nu_{\pi^+}=1.8$ is outside the range, and we use $R^\nu_{\pi^+}=1.6$ instead. It is observed that the contours of these observables in the $y^R_{T3}$-$y^L_{T3}$ plane exhibit distinct patterns. $R^\nu_{K^*}$ and ${\cal B}(B_s \to \mu^- \mu^+)$ behave like hyperbolic curves, while $R^\nu_{K}$ and $R^\nu_{\pi}$ exhibit circular patterns.  In Fig.~\ref{fig:obsyR3yR2}(b), the $R^\nu_{\pi^0}=1.2$ and $1.4$ contours do not overlap with the parameter region within the $3\sigma$ CL; however, the $R^\nu_{\pi^0}=1.2$ and $1.4$ contours do cross over the region when viewed in the $y^R_{T3}$-$y^L_{T3}$ plane of Fig.~\ref{fig:obsyR3yR2}(d).

According to the results in Fig.~\ref{fig:obsyR3yR2}, we observe that when $y^{R(L)}_{T3}\sim y^{R(L), \rm min}_{T3}$, $R^\nu_{K, \pi^+,\pi^0}$ can be significantly enhanced and that ${\cal B}(B_s\to \mu^-\mu^+)$ can fit the experimental central value. Thus, it is interesting to explore the dependence of the considered processes on the $y^{R,L}_{T2}$ parameters, with the other parameter values fixed at ${\bf y}^{R,\rm min}_T$ and ${\bf y}^{L, \rm mini}_T$. For illustration purposes, we show the contours as functions of $y^{L}_{T2}$ and $y^{R}_{T2}$ for $R^\nu_{K, K^*}$ and ${\cal B}(B_s \to \mu^- \mu^+)$  in Fig.~\ref{fig:yL2yR2}(a) and for $R^{\nu}_{\pi^+, \pi^0}$ in Fig.~\ref{fig:yL2yR2}(b). From the plots, it is evident that each observable in the selected values of contours can match the area within the $3\sigma$ CL. However, from Fig.~\ref{fig:yL2yR2}(a), the curves of $R^\nu_{K^*}=0.9$ and $1.2$ do not overlap with the contours of $B_s\to \mu^- \mu^+$ in the $R^\nu_K$-enhanced region within the $1\sigma$ error. The same behavior is also shown in Fig.~\ref{fig:obsyR3yR2}(c). Hence, $R^\nu_{K^*}$ can be strictly bound by the measurement of ${\cal B}(B_s \to \mu^- \mu^+)$.

\begin{figure}[phtb]
\begin{center}
\includegraphics[scale=0.4]{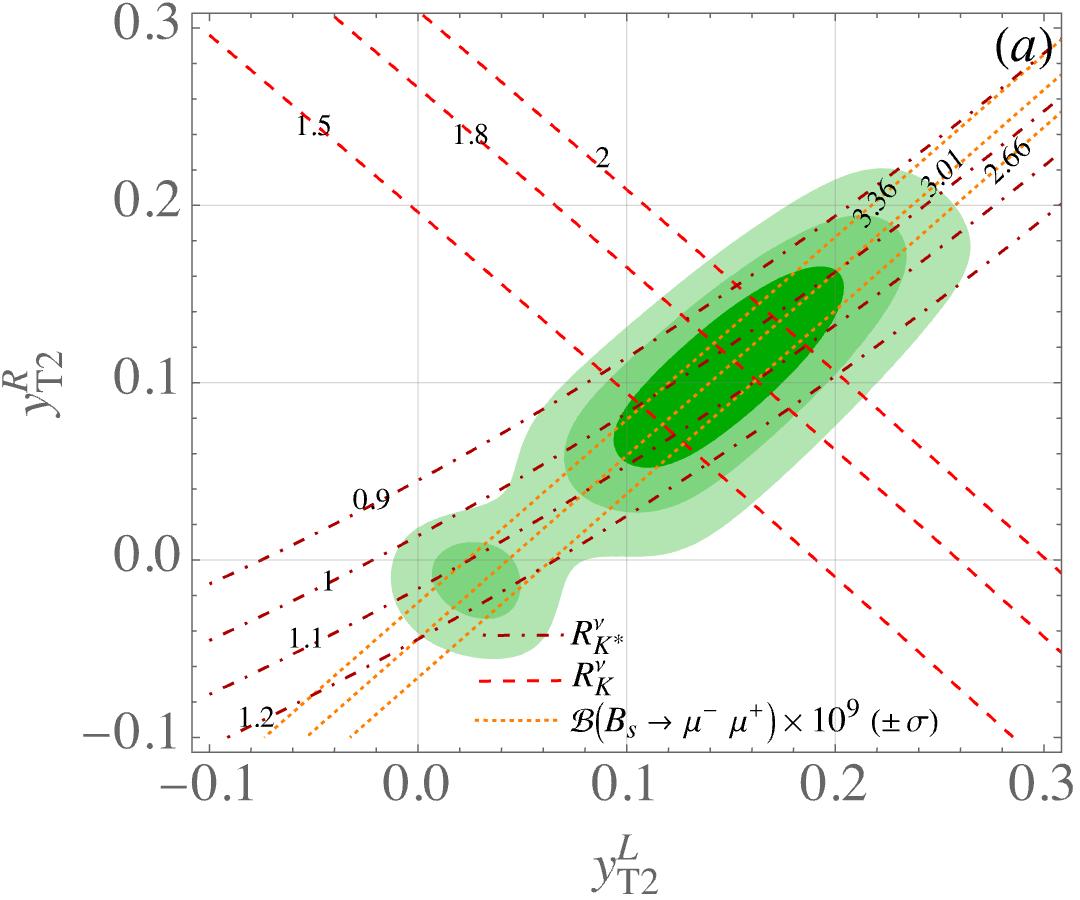}\hspace{3mm}
\includegraphics[scale=0.4]{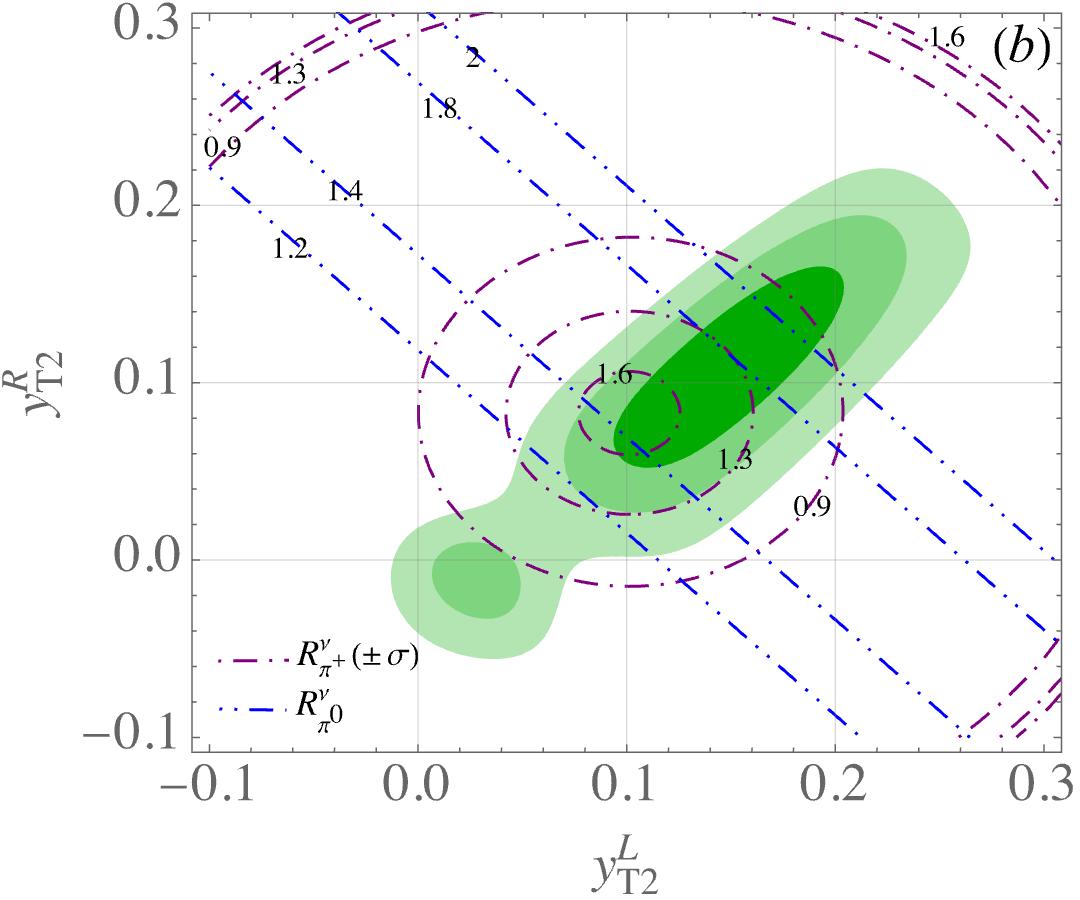}
 \caption{ Contours in the $y^L_{T2}$-$y^R_{T2}$ plane for the observables of (a) $R^\nu$, $R^\nu_{K^*}$, and ${\cal B}(B_s\to \mu^- \mu^+)$ and of (b) $R^\nu_{\pi^+}$ and $R^\nu_{\pi^0}$, where the parameter values, except for $y^{L,R}_{T2}$, are set to be ${\bf y}^{R(L),\rm min}_{T}$. In the plots, $\pm \sigma$ denotes that the experimental data within one standard deviation are applied. }
\label{fig:yL2yR2}
\end{center}
\end{figure}

To examine the correlations between observables, we need to vary all parameters simultaneously instead of merely varying two parameters as before.  Based on the above analysis, except for $y^{R(L)}_{T1}$, which are fixed at the values of $y^{R(L), \rm min}_{T1}$, we set the ranges of the other parameters around ${\bf y}^{R(L),\rm min}_{T}$.  We scan the parameters assuming a normal distribution with ${\bf y}^{R(L),\rm min}_{T}$ as the mean and setting the standard deviations as $\sigma(y^{R(L)}_{T2})=y^{R(L),\rm min}_{T2}$ and $\sigma(y^{R(L)}_{T3})=0.1$. In addition to the $1\sigma$ experimental constraints shown in Table~\ref{tab:inputs}, we further require that $\chi^2-\chi^2_{\rm min} \lesssim 3.84$.

Using $10^7$ sampling points, Fig.~\ref{fig:scan} shows the scatter plot depicting the correlations between $R^\nu_K$ and $R^\nu_M$ with $M=K^*, \pi^+, \pi^0$ at 95\% CL. As alluded to before, the result of $R^\nu_K\simeq R^\nu_{\pi^0}$ given in Eq.~(\ref{eq:RnuK=Rnupi0}) is numerically verified. Both ratios can be enhanced up to a factor of 2. As expected, due to the introduction of right-handed quark currents, which are used to fit the observed BR for $B_s\to \mu^- \mu^+$,  the influence of new physics effects on $R^\nu_{K^*}$ in the model is mild. It is observed that both $R^\nu_{\pi^+}$ and $R^\nu_K$ increase simultaneously when $R^\nu_K \lesssim 1.4$. Subsequently, when $R^\nu_K > 1.4$, $R^\nu_{\pi^+}$ decreases as $R^\nu_K$ increases. This behavior of $R^\nu_{\pi^+}$ can be understood as follows.  The imaginary part of $X_{\rm eff}$, which also contributes to $K_L\to \pi^0 \nu \bar\nu$ and is defined in Eq.~(\ref{eq:xeff}), increases linearly with $R^\nu_K$. However, the real part of $X_{\rm eff}$ decreases after $R^\nu_K \simeq 1.4$. Since $K^+\to \pi^+ \nu \bar\nu$ is dominated by the CP-conserving effect, this leads to a decrease in ${\cal B}(K^+\to \pi^+ \nu \bar\nu)$ for $R^\nu_K > 1.4$, although the BR of the CP-violating process $K_L\to \pi^0 \nu \bar\nu$ continues to increase.  Since the resulting ${\cal B}(B_s \to \mu^- \mu^+)$ can be within $1\sigma$ errors of experimental data for any value of  $R^\nu_K$ in the region of  $(1.2, \, 2.2)$,  the scatter plot for the correlation between ${\cal B}(B_s \to \mu^- \mu^+)$ and $R^\nu_K$ is not shown.  In all interesting regions of $R^\nu_K$, the $Z$-penguin diagrams mediated by the charged Higgs bosons can alleviate the tension observed in $B_s \to \mu^- \mu^+$ between the experimental measurement and the SM prediction.

\begin{figure}[phtb]
\begin{center}
\includegraphics[scale=0.6]{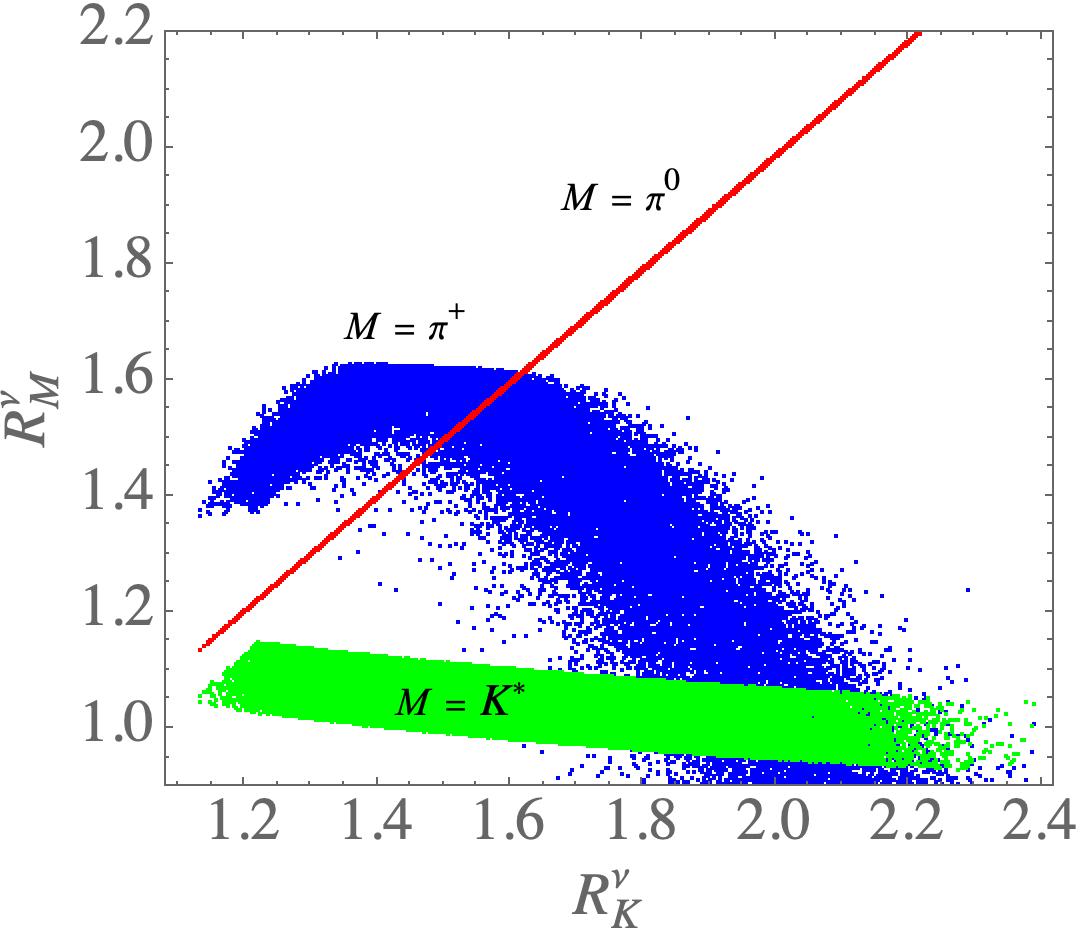}
 \caption{Scatter plot illustrating the correlations between  $R^\nu_K$ and  $R^\nu_M$ with $M=K^*, \pi^+$, and $\pi^0$.}
\label{fig:scan}
\end{center}
\end{figure}

\section{Summary} \label{sec:sum}

Scotogenesis, typically used for lepton-related phenomena such as neutrino mass and dark matter candidate, is rarely studied in the context of the quark flavor-related processes. Moreover, the minimal scotogenic model~\cite{Ma:2006km} cannot accommodate the muon $g-2$ anomaly. To apply the mechanisms in a scotogenic model to the quark FCNC processes and preserve the characteristic features of scotogenesis, we propose in this work a scotogenic model whose dark particles include a neutral Dirac-type dark lepton, two inert Higgs doublets with one carrying a dark charge, a charged singlet dark scalar, and a singlet vector-like up-type dark quark. As demonstrated above, each of these particles plays a crucial role in this study.

With appropriate dark charge assignments, the model has a dark $U(1)$ symmetry with which the scalar and pseudoscalar in the same inert Higgs doublet are degenerate in mass. Consequently, the inert scalars cannot be dark matter candidates. Nevertheless, the singlet dark neutral lepton does not couple to the $Z$ boson and the SM Higgs boson, and can self-scatter into the SM particles through the Yukawa couplings; thus, it can be dark matter in the model.

With the introduction of a dark up-type quark, the SM quarks can couple to this dark quark via the non-leptonic inert Higgs doublet. As a result, the loop-induced $Z$-penguin diagrams make significant contributions to the $B^+\to K^+ \nu \bar\nu$ and $K\to \pi \nu \bar\nu$ processes. Furthermore, with the introduction of a singlet dark-charged scalar, not only can muon $g-2$ be enhanced, but also the right-handed quark current for the $b\to s$ transition can be induced. As a result, the tension in $B_s\to \mu^- \mu^+$ between the experimental measurement and the SM prediction can be resolved.

\section*{Acknowledgments}
C.-W.~C would like to thank the High Energy Physics Theory Group at the University of Tokyo for their hospitality during his visit when part of this work was finished.
This work was supported in part by the National Science and Technology Council, Taiwan, under Grant Nos. NSTC-110-2112-M-006-010-MY2 (C.-H.~Chen) and NSTC-111-2112-M-002-018-MY3 (C.-W.~Chiang).

\appendix

\section{Saclar potential and masses of scalars} \label{app:potential}

The scalar potential for the SM Higgs $H$, $\eta_{1,2}$, and $\chi^+$ is written as:
\begin{align}
\begin{split}
 V =& V_{\rm SM} + V(H,\eta_1,\eta_2,\chi^\pm)~, 
 \\
 V_{\rm SM} =& - \mu^2_H H^\dag H + \lambda_H (H^\dag H)^2 ~,
 \\
 V(H, \eta_1,\eta_2,\chi^\pm ) =& 
 \mu^2_1 \eta^\dag_1 \eta_1+ \mu^2_2 \eta^\dag_2 \eta_2 + \mu^2_3 \chi^- \chi^+ + (\mu_\chi \eta^T_1 i\tau_2 H \chi^-  + \mbox{H.c.}) + \lambda_1 (\eta^\dag_1 \eta_1)^2 
 \\
 &+ \lambda_2 (\eta^\dag_2 \eta_2)^2  + \lambda_3 (\eta^\dag_1 \eta_1) (\eta^\dag_2 \eta_2)+ \lambda_4 (\eta^\dag_1 \eta_2) (\eta^\dag_2 \eta_1)  
 \\ 
& + \left( \lambda_5 (H^\dag \eta_1) (H^\dag \eta_2) + \mbox{H.c.}\right) 
 + \lambda_6 (H^\dag \eta_1) (\eta^\dag_1 H) + \lambda_7  (H^\dag \eta_2) (\eta^\dag_2 H)  
 \\
 &+  \lambda_8 (H^\dag H) (\eta^\dag_1 \eta_1) +  \lambda_9 (H^\dag H) (\eta^\dag_2 \eta_2) + \lambda_{10} (\chi^- \chi^+)^2  \\
 &+   (\chi^- \chi^+) (\lambda_{11} H^\dagger H + \lambda_{12} \eta^\dag_1 \eta_1 + \lambda_{13} \eta^\dag_2 \eta_2) ~\label{eq:scalar_potential}.
\end{split}
\end{align}
It can be seen that the non-self-Hermitian terms are the $\lambda_5$ and $\mu_\chi$ terms, where the former violates the lepton number by two units and plays an important role in the radiative generation of the Majorana neutrino mass, and the latter leads to the mixing between $\eta^\pm_1$ and $\chi^\pm$.  In addition to the leptonic Yukawa couplings shown in Eq.~(\ref{eq:yukawa}), the tiny neutrino mass can be achieved when $\lambda_5 \ll 1$.  
To spontaneously break the electroweak gauge symmetry, we take $\mu^2_H, \lambda_H >0$ as in the SM.  The masses of $\eta_{1,2}$ can be irrelevant to the electroweak symmetry breaking, and we thus require $\mu^2_{1,2}(\lambda_{1,2}) > 0$.  Using the charge assignment shown in Eq.~\eqref{eq:rep}, the scalar potential in Eq.~(\ref{eq:scalar_potential}) has an unbroken global $U(1)_X$  symmetry. To preserve the global $U(1)_X$ symmetry,  the VEVs of $\eta_{1,2}$ are kept zero.  Therefore,
the components of the three doublet scalars are parametrized as follows ($j = 1,2$): 
\begin{equation}
  H= 
\left(
\begin{array}{c}
  G^+     \\
  \frac{1}{\sqrt{2}} ( v + h + i G^0)      \\   
\end{array}
\right)~, ~~  \eta_j= 
\left(
\begin{array}{c}
  \eta^+_j    \\
  \frac{1}{\sqrt{2}} (\eta^0_j + i a_j)    \\   
\end{array}
\right)~, \label{eq:H_eta_rep}
\end{equation}
where $G^{\pm,0}$ are the Goldstone bosons, $v$ is the VEV of $H$, and  $h$ is the SM Higgs boson.

From Eq.~(\ref{eq:scalar_potential}), $\eta^\pm_2$ does not mix with the other charged scalars.  Its mass is given by:
 \begin{equation}
 m^2_{\eta^\pm_2}  = \mu^2_2 + \frac{\lambda_9 v^2}{2}~.
 \end{equation}
Because $\eta^\pm_1$ mixes with $\chi^\pm$ via the $\mu_\chi$ term, their mass-squared matrix is found to be:
\begin{align}
& (\eta^-_1, \chi^-)   
\left(
\begin{array}{cc}
 m^2_{11} &   m^2_{12}   \\
m^2_{12}  &  m^2_{22}    \\    
\end{array}
\right)  \left(
\begin{array}{c}
 \eta^+_1  \\
  \chi^+      \\    
\end{array}
\right)\,, \nonumber \\
& \mbox{with}~
m^2_{11} = \mu^2_1 + \frac{\lambda_8 v^2}{2}\,, ~m^2_{12} = \frac{\mu_\chi v}{\sqrt{2}}\,, ~ m^2_{22}  = \mu^2_3+ \frac{\lambda_{11} v^2}{2}\,.
\end{align}
The mass matrix can be diagonalized by a $2\times 2$ orthogonal rotation defined by
\begin{equation}
\left(
\begin{array}{c}
 H^+_1  \\
H^+_2      \\    
\end{array}
\right) = \left( \begin{array}{cc}
c_{\theta}  & s_{\theta} \\
-s_{\theta}     &  c_{\theta} \\    
\end{array}
\right)  \left(
\begin{array}{c}
 \eta^+_1   \\
 \chi^+      \\    
\end{array}
\right)\,,  \label{eq:mixing}
\end{equation}
where $c_{\theta} \equiv \cos\theta$ and $s_{\theta} \equiv \sin\theta$. As a result,  the mass eigenvalues and the mixing angle are obtained as:
\begin{align}
  m^2_{1(2)} &= \frac{m^2_{11} + m^2_{22}}{2} \pm \frac{1}{2} \sqrt{(m^2_{22} - m^2_{11})^2 + 4 (m^2_{12})^2}
  \,, \nonumber \\
  s_{2\theta} & = - \frac{2 m^2_{12} }{m^2_{2} - m^2_{1} }\,,  \label{eq:theta_chi}
\end{align}
where $s_{2\theta} \equiv \sin2\theta$.

Since the terms $(H^\dag \eta_j)^2$ are forbidden by the $U(1)_X$ symmetry, the scalar $\eta^0_j$ and pseudoscalar  $a_j$ are degenerate in mass. Although the $\lambda_5$ term would mix $\eta^0_1(a_1)$ and $\eta^0_2(a_2)$, it will not lift the degeneracy. To obtain the mass spectrum of the neutral scalar bosons, we write the mass matrices for $(\eta^0_1, \eta^0_2)$ and $(a_1,a_2)$ as follows:
 \begin{equation}
m^2_{S} = \left(
\begin{array}{cc}
  \bar m^2_{11} &   \bar m^2_{12}  \\
 \bar m^2_{12}   &   \bar m^2_{22} \\   
\end{array}
\right)~, ~~ m^2_{A} =\left(
\begin{array}{cc}
  \bar m^2_{11} &    - \bar m^2_{12}  \\
 - \bar m^2_{12}   &   \bar m^2_{22} \\   
\end{array}
\right)~, 
 \end{equation}
where the matrix elements are:
\begin{align}
 \bar m^2_{11} & = \mu^2_{1} + \frac{v^2}{2} (\lambda_6 + \lambda_8) 
 ~, \nonumber \\
\bar m^2_{22} & = \mu^2_2 + \frac{v^2}{2} \left( \lambda_7 + \lambda_9 \right)
~, \nonumber \\ 
 \bar m^2_{12} & = \frac{v^2}{2} \lambda_5 ~.
\end{align}
Analogous to the case of the charged scalar mass matrix, each of the two $2\times 2$ mass-squared matrices can be diagonalized by the corresponding orthogonal matrix $O(\theta_\zeta)$ ($\zeta= S, A$) through $O(\theta_\zeta) m^2_{\zeta} O^T(\theta_\zeta)$, where
\begin{equation}
  O(\theta_\zeta) = \left(
\begin{array}{cc}
 \cos\theta_\zeta &   \sin\theta_\zeta  \\
 -\sin\theta_\zeta &   \cos\theta_\zeta \\   
\end{array}
\right)~. \label{eq:omatrix}
\end{equation}
Since the matrix elements in $m^2_A$ are the same as those in $m^2_S$ except for the sign change in the off-diagonal elements, we therefore take $\theta_S=- \theta_A \equiv\phi$.  The eigenvalues of $m^2_S$ are found to be:
\begin{equation}
m^2_{S_{1,2}} = \frac{1}{2} \left[ \bar m^2_{11} + \bar m^2_{22} \pm \sqrt{(\bar m^2_{22} - \bar m^2_{11} )^2 - 4 (\bar m^4_{12})} \right]~.
\end{equation}
For the physical pseudoscalars $A_{1,2}$, we have $m^2_{A_{1(2)}}= m^2_{S_{1(2)}}$.  The mixing angle $\phi$ is defined by:
\begin{equation}
\sin2\phi = - \frac{\lambda_5 v^2}{m^2_{S_2} - m^2_{S_1}}~.
\end{equation}
Since the $\lambda_5$ term violates the lepton number and eventually leads to the Majorana mass, its value has to be sufficiently small, i.e., $\lambda_5 \ll 1$.  As a result, the off-diagonal mass matrix element $|\bar m_{12}^2|$ is suppressed and the mixing angle $\phi \ll 1$.

\section{Form factors for $\bar B\to (K, K^*)$} \label{app:FFs}

The $q^2$-dependent form factors for $\bar B\to K$ are defined through the following relations:
\begin{align}
\langle K(p_{2} )| V_{\mu }| \bar{B} (p_{1})\rangle &=
f_{+}(q^2)\left( P_{\mu}-\frac{P\cdot q }{q^2}\right)
+\frac{P\cdot q}{q^2}f_{0}(q^2)\,q_{\mu},  \nonumber \\
\langle K(p_{2} )| T_{\mu\nu }q^{\nu}| \bar{B}
(p_{1})\rangle &= {f_{T}(q^2)\over m_{B}+m_{K}}\left(P\cdot q\,
q_{\mu}-q^{2}P_{\mu}\right)~,
\end{align}
where $P=p_1+p_2$, $q=p_1 - p_2$; $V_{\mu }=\bar{s}\gamma _{\mu } b$, and $T_{\mu \nu }=\bar{s} i\sigma _{\mu \nu }b$.  For $\bar B\to K^*$, the form factors are parametrized as:
\begin{align}
\begin{split}
\langle K^{*}(p_{2},\epsilon )| V_{\mu }| \bar{B} (p_{1})\rangle 
=&
i\frac{V(q^{2})}{m_{B}+m_{K^{*}}}\varepsilon
_{\mu \alpha \beta \rho }\epsilon ^{*\alpha }P^{\beta }q^{\rho }
~, \\
\langle K^{*}(p_{2},\epsilon )| A_{\mu }| \bar{B}
(p_{1})\rangle 
=&
2m_{K^{*}}A_{0}(q^{2})\frac{\epsilon ^{*}\cdot q}{
q^{2}}q_{\mu }+( m_{B}+m_{K^{*}}) A_{1}(q^{2})\left( \epsilon
_{\mu }^{*}-\frac{\epsilon ^{*}\cdot q}{q^{2}}q_{\mu }\right)  \\
&-A_{2}(q^{2})\frac{\epsilon ^{*}\cdot q}{m_{B}+m_{K^{*}}}\left( P_{\mu }-
\frac{P\cdot q}{q^{2}}q_{\mu }\right) 
~,  \\
\langle K^{*}(p_{2},\epsilon )| T_{\mu \nu }q^{\nu }| \bar{B}
(p_{1})\rangle 
=& -iT_{1}(q^{2})\varepsilon _{\mu \alpha \beta
\rho
}\epsilon ^{*\alpha }P^{\beta }q^{\rho }
~,  \\
\langle K^{*}(p_{2},\epsilon )| T_{\mu \nu }^{5}q^{\nu }|
\bar{B}(p_{1})\rangle 
=& T_{2}(q^{2})\left( \epsilon _{\mu
}^{*}P\cdot q-\epsilon ^{*}\cdot qP_{\mu }\right)
+T_{3}(q^{2})\epsilon ^{*}\cdot q\left( q_{\mu
}-\frac{q^{2}}{P\cdot q}P_{\mu }\right) ~.\label{ffv}
\end{split}
\end{align}
Here $\epsilon$ denotes the polarization vector of $K^*$, $A_{\mu }=\bar{s} \gamma_{\mu}\gamma _{5} b$, and $T_{\mu \nu }^{5}=\bar{s} i\sigma _{\mu \nu }\gamma _{5} b$. It is useful to define additional form factors $A_{12}(q^2)$ and $T_{23}(q^2)$:
 \begin{align}
 16 m_B m^2_{K^*} A_{12}(q^2) &= (m_B + m_{K^*}) (m^2_B - m^2_{K^*} -q^2) A_1 (q^2) -\frac{ \lambda_{K^*}(q^2) }{m_B + m_{K^*}}A_2(q^2) ~, \\
 8 m_B m^2_{K^*} T_{23}(q^2) & =  (m_B+ m_{K^*}) (m^2_B + 3 m^2_{K^*} -q^2 ) T_2(q^2)- \frac{\lambda_{K^*}(q^2)}{m_B - m_{K^*}} T_3(q^2)~. 
 \end{align}


\end{document}